\documentclass[sigconf,screen,prologue,table,xcdraw]{acmart}

\usepackage{tikz}
\usetikzlibrary{positioning, arrows.meta}
\usepackage{subcaption}
\usepackage{multirow}
\usepackage{booktabs}
\usepackage{array}
\usepackage{ragged2e}
\usepackage{lipsum}
\usepackage{caption}
\usepackage{longtable}
\usepackage{listings}

\usepackage{tabularx}
\usepackage{float}
\usepackage{amsmath}

\definecolor{color2}{RGB}{64,64,64}

\newif\ifshowchanges
\showchangestrue
\showchangesfalse  

\ifshowchanges
    \usepackage[commandnameprefix=ifneeded]{changes}
\else 
    \newcommand{\added}[1]{#1}
    \newcommand{\deleted}[2]{}
\fi

\AtBeginDocument{%
  }

\setcopyright{none}
\begin{document}
\title[Tradeoffs in LLM-Supported Research Ideation]{Who Owns Creativity and Who Does the Work? Trade-offs in LLM-Supported Research Ideation}

\author{Houjiang Liu}
\authornote{Both authors contributed equally as first authors to this research.}
\affiliation{%
  \institution{School of Information, University of Texas at Austin}
  \city{Austin}
  \state{Texas}
  \country{USA}
}

\author{Yujin Choi}
\authornotemark[1]
\affiliation{%
  \institution{School of Information, University of Texas at Austin}
  \city{Austin}
  \state{Texas}
  \country{USA}
}

\author{Sanjana Gautam}
\affiliation{%
  \institution{School of Information, University of Texas at Austin}
  \city{Austin}
  \state{Texas}
  \country{USA}
}

\author{Gabriel Jaffe}
\affiliation{%
  \institution{Texas Advanced Computing Center}
  \city{Austin}
  \state{Texas}
  \country{USA}
}

\author{Soo Young Rieh}
\authornote{Both authors contributed equally as corresponding authors to this research.}
\affiliation{%
  \institution{School of Information, University of Texas at Austin}
  \city{Austin}
  \state{Texas}
  \country{USA}
}

\author{Matthew Lease}
\authornotemark[2]
\affiliation{%
  \institution{School of Information, University of Texas at Austin}
  \city{Austin}
  \state{Texas}
  \country{USA}
}

\renewcommand{\shortauthors}{Houjiang Liu et al.}

\begin{abstract}
LLM-based agents offer new potential to accelerate science and reshape research work. However, the quality of researcher contributions can vary significantly depending on human ability to steer agent behaviors. How can we best use these tools to augment scientific creativity without undermining aspects of contribution and ownership that drive research? To investigate this, we developed an agentic research ideation system integrating three roles---Ideator, Writer, and Evaluator---across three control levels---Low, Medium, and Intensive. Our mixed-methods study with 54 researchers suggests three key findings in how LLM-based agents reshape scientific creativity: 1) perceived creativity support does not simply increase linearly with greater control; 2) human effort shifts from ideating to verifying ideas; and 3) ownership becomes a negotiated outcome between human and AI. Our findings suggest that LLM agent design should emphasize researcher empowerment, fostering a sense of ownership over strong ideas rather than reducing researchers to operating an automated AI-driven process.

\end{abstract}



\settopmatter{printfolios=true}


\maketitle

\section{Introduction} \label{sec:intro}


Research ideation in science is a human-centered creative process that requires deep domain expertise \citep{Stremersch2024-wl}, methodological judgment \citep{Si2025-zr}, and sustained scientific reasoning \citep{Glass2008-sk, Schickore2025-on}. It involves not only generating novel ideas but also evaluating, refining, and aligning them with disciplinary norms and long-term research goals \citep{Kuhn2012-zv}. The rise of large language models (LLMs) offers new potential to assist and/or automate research tasks traditionally requiring human effort, such as ideation \citep{Wang2023-hl, Zhang2025-uk}, annotation \citep{Tan2024-ll}, evaluation \citep{Bavaresco2025-yj, Li2024-ov}, and even scientific prediction \citep{Luo2025-ji, Luo2025-bv}. 

In regard to LLM-supported research ideation, recent studies report mixed results, especially with agentic systems designed to create a coherent, end-to-end idea elaboration workflow that spans seed idea generation, literature synthesis, and the formulation of research hypotheses and study plans. While some prior work shows that these agent-based ideation tools can help researchers better use their expertise to produce more novel ideas \citep{Radensky2024-gq, Si2024-nm}, other work reports that such systems only offer short-term boosts while hindering deeper cognitive engagement during the creative process \citep{Kumar2025-og}, or that the resulting ideas lack real-world feasibility \citep{Si2025-zr}. 

One potential source of these differences is that agent-based tools differ in \textit{steerability} \cite{Feng2024-xl}: the degree of control researchers can guide, constrain, and iteratively refine agent behaviors, resulting in potential trade-offs between LLM creative assistance and human effort. 
%
A key consequence of steerability is how it impacts a researcher sense of ownership---closely tied to scientific accountability and responsibility---over research ideas generated with AI support. Researchers must defend, validate, and ethically stand behind their ideas, a burden that AI cannot shoulder \citep{Bozkurt2024-ds}. Currently, it remains unclear how the creative labor division between humans and AI shapes the trade-off between creativity support and effort, and how this division influences researcher sense of ownership during the iterative development of research ideas. Better understanding such tradeoffs would help better inform researchers considering use of such tools and in choosing between alternatives with varying degrees of steerability. Designers today also
lack insights into designing systems that balance control and constraints to optimize user experience while ensuring fair, transparent, and legally informed attribution of credit and responsibility.

In this work, we explore two key research questions:
\begin{enumerate}
\item How do different levels of control over LLM agents affect the potential trade-offs between creativity support and the effort required for research ideation? 
\item How do these control levels influence the division of contributions between humans and AI, as well as researcher perceived ownership of co-created research ideas?
\end{enumerate}
To investigate these questions, we conceptualize research ideation as a form of scientific creativity, reflecting the intellectual effort researchers invest in advancing new knowledge. We develop an agentic system that provides an end-to-end idea elaboration process for research proposal writing. The system integrates three agentic roles---Ideator, Writer, and Evaluator --- with controls for different proposal stages. These controls refer to user-initiated interactions throughout idea elaboration process, including generating seed ideas, specifying literature focus, revising proposal content, and giving feedback on evaluation results and proposal improvements. We further differentiate these controls into three human supervised levels: Low, Medium, and Intensive. Higher levels provide users with increasingly fine-grained interaction capabilities. Separating these levels serves as a set of design probes to explore trade-offs and encourage researcher reflection on appropriate tool use:
\begin{description}
    \item[Low] This level reflects a minimally supervised agentic workflow, where humans initiate the process, after which idea elaboration proceeds entirely by agents. Researchers provide initial keywords to generate and customize seed ideas, and the system automatically determines relevant literature, generates proposal content, and self-evaluate proposal for improvements without user involvement.
    \item[Medium] This level reflects a partially supervised agentic workflow, where humans can steer high-level direction and adjust proposal focus, while agents continue to elaborate idea details. Researchers can select the literature focus and adjust proposal sections through prompting, but they have limited ability to make fine-grained revisions, particularly when it comes to editing generated content, giving detailed feedback on evaluation results, or guiding future improvements for a complete revision.
    \item[Intensive] This level reflects a fully supervised agentic workflow, where humans maintain complete control over both high-level direction and fine-grained content development, with agents assisting iterative refinement. Researchers take full control over generating and expanding seed ideas, including specifying literature focus, iteratively revising proposal content through editing and inline-text prompting, evaluating agent self-feedback, or customizing improvements.
\end{description}

We conducted a mixed-methods study that invited researchers to use the three control levels (between-subjects), followed by semi-structured interviews. Both quantitative and qualitative data were collected. Quantitative measures included behavioral metrics, self-reported creativity support, cognitive effort, and perceived ownership, while qualitative insights were participant reflections after tool usage. Our 54 researcher participants represented diverse disciplines across the natural sciences and engineering, computer and information sciences, and social sciences and humanities.

Our results reveal a complex landscape of trade-offs in the division of creative labor, with final ownership emerging as a negotiated outcome between researchers and AI. Quantitative data shows that perceived creativity support does not simply increase linearly with greater control. Instead, Low- and Intensive-control levels foster distinct types of creative gains, each associated with different cognitive effort. Qualitative analysis further reveals a shift in human labor. When using the tool felt low-effort for generating ideas, especially in the low-control condition, this came with mixed-quality outputs and reduced cognitive engagement in proposal development, a pattern consistent across control levels. At the same time, participants reported spending more effort on verifying ideas and making sense of the system.

Finally, we observed distinct quantitative patterns in how participants attributed contributions between humans and AI across control levels. Those who classified the proposal as either `Human work' or `AI work' consistently viewed one side as the primary contributor, regardless of control levels. In contrast, `Co-Created Work' showed more varied patterns across control levels. Qualitative data suggests these differences stemmed from how researchers valued the originality of ideas, the execution effort involved in proposal development, and their perceptions of who held control.


\textbf{Contributions.} Our study investigates the trade-offs between creativity support and effort across three control levels to steer LLM agents for research ideation, and how these different levels influence researcher perceptions of ownership over final co-created proposals. 
The complex trade-offs between creativity support and effort, and perceived ownership underscore the importance of tailoring LLM controls to diverse researcher goals. We conclude that designing LLM agents for research ideation should empower scientists, allowing them to experience a greater sense of ownership over more powerful ideas, rather than reducing them to operators of a machine that controls the creative process. 


\section{Related Work} \label{sec:lit-review}

\subsection{Research Ideation in Scientific Creativity} \label{subsec:scientific_creativity}

Research ideation is a distinctive form of creative practice and a central activity through which scientific creativity is investigated. From a philosophical and historical perspective, scientific creativity is characterized by a researcher ability to form a preliminary conjecture about an underlying problem and a potential solution, or, more broadly, to generate an interesting idea to advance domain knowledge \citep{Sawyer2023-ud, Stumpf1995-wa}. As first theorized by 19th-century scholar William Whewell in his model of scientific discovery \citep{Whewell1840-hw}, an idea is ``a happy thought'' from researchers, followed by the ``articulation'' and ``verification'' of that thought \citep{Schickore2025-on}. In recent decades, creativity scholars, such as \citet{Sawyer2024-mm, Simonton2004-iu, Boden2003-sr}, have increasingly studied research ideation as a distinct mode of creative practice emerging from the processes of scientific discovery. People everyday or artistic work, such as visual arts \citep{Sawyer2023-fw}, creative writing \citep{Sawyer2023-jk}, and other activities performed by painters, musicians, or writers, draws primarily from the imaginative and expressive aspects of human creativity. In comparison, research ideation represents reasoning modes characteristic of scientific work, including deductive, inductive, and analogical thinking, all grounded in theory and experimentation \citep{Sawyer2023-ud}.

The complexity of research ideation not only arises from its distinctive nature as a part of scientific discovery, as described above, but also from the diversity of its sub-activities and their underlying cognitive processes. Under the epistemic position of ``Cognitive Emergence,'' synthesized in a theoretical review of creativity research in HCI by \citet{Hsueh2024-vv}, research ideation can be conceptualized as a multi-phase human creative reasoning process. For example, mapping research ideation to \citet{Wallas1926-em}'s classic model reveals distinct sub-activities, such as sensemaking and gap identification in the \textit{preparation} phase, brainstorming research ideas such as hypotheses or potential solutions in \textit{incubation} and \textit{illumination} phase, and refining ideas during \textit{verification} phase.

The above ideation process is inherently demanding, requiring deep domain expertise and sustained cognitive engagement. From a human-centered perspective, the effort researchers invest is not merely a burden but a fundamental driver of creativity \citep{Kumar2025-og}. As highlighted by \citet{Csikszentmihalyi1990-wm}'s {\em Flow Theory}, the challenge of reconciling novelty with rigor gives rise to the ``Optimal Experience'' of a creative expert. \citet{Sawyer2023-ud} characterize this as moments of elegance, wonder, or frustration integral to meaningful research ideation. Bearing these cognitive insights in mind, HCI scholars have long explored creativity support tools that go beyond the efficiency gains of ideas, recognizing and preserving researcher cognitive flow central to idea quality \citep{Hsueh2024-vv}.

\subsection{Design Spectrum for LLM-Supported Research Ideation} \label{subsec:design_spectrum}

As LLMs and agentic systems gain prominence, there has been growing momentum to develop tools that support research ideation. Broadly, these systems can be characterized into two design patterns: automated systems and mixed-initiative tools. This distinction aligns with \citet{Shneiderman2020-yb}'s summary of historical AI design goals: (1) the emulation goal, aiming to build autonomous and intelligent agents that can perform tasks as well as or even better than humans, and (2) the application goal, aiming to create tools that augment human capabilities and needs, and enable effective human-AI collaboration.

While automated capabilities extend beyond ideation to span the broader research process, such as experimental execution, data visualization, and writing reports, ideation remains the critical first step in developing an end-to-end scientific discovery system that emulates human researchers \citep{Sourati2023-ca, Baek2025-ew}. For example, \textit{SciAgents} \cite{Ghafarollahi2024-tl} generates potential bio-material ideas using ontological knowledge graphs grounded in materials science concepts. The \textit{AI Scientist} \cite{Lu2024-jg} is an autonomous system capable of conducting computer science research and producing complete research papers. Similarly, \textit{Agent Laboratory} \citep{Schmidgall2025-ff} and \textit{AI Co-Scientists} \citep{Gottweis2025-hh} can perform large-scale, multi-turn exploration to screen and generate novel research ideas across scientific domains.

HCI scholars have long explored mixed-initiative design to support human creativity in different domains (in our case, research ideation) \citep{Horvitz1999-jx, Hsueh2024-vv}, where humans and AI can each take the initiative on different parts of a creative task, thereby leveraging the complementary strengths of both. Reflecting on the phase-based ideation process (like \citet{Wallas1926-em}, described in Section \ref{subsec:scientific_creativity}), a wide range of tools have been developed to support separated sub-activities of research ideation. For example, in the incubation phase, \citet{Kang2023-yh} developed \textit{Synergi}, which allows researchers to explore thread-based research concepts relevant to their interests based on citation graphs and LLM summarization. \citet{Liu2024-he} created \textit{CoQuest} for idea incubation, which helps researchers to brainstorm research questions with LLMs using either a breadth-first or depth-first prompting strategy. In the verification phase, \citet{Radensky2024-gq} built \textit{Scideator} that lets researchers to construct criteria facets then scan and filter novel ideas with cross-domain insights.

As prior mixed-initiative tools were designed to support only discrete stages of research ideation, the rise of automated pipelines powered by LLM agents has prompted scholars to explore more coherent, end-to-end ideation workflows. This trend reflects increasing convergence between the AI and HCI communities, where humans and agents jointly contribute across the full ideation pipeline. Within this shift, proposal writing recently emerged as a key study context, where tasks such as literature synthesis, gap identification, hypothesis formation, and study planning are tightly interconnected, aiming to transform an initial, undeveloped idea into a concrete draft \citep{Li2025-yj}. For example, \citet{Pu2025-us} developed \textit{IdeaSynth}, which supports searching, summarizing, and drafting research proposals on a canvas-based ideation board. \citet{Garikaparthi2025-fs} developed \textit{IRIS}, enabling researchers to engage in a similar workflow through a chat interface.

As human–AI integration deepens, such as moving from isolated activities toward continuous, cross-activity workflow, it becomes critical to understand how work should be divided between humans and AI. This shift raises potential tensions in role allocation and how to design for meaningful, productive idea elaboration process.

\subsection{Balancing User Control and Automation in LLM-Supported Research Ideation} \label{subsec:steerability_influence}

Building on the earlier discussion of automated solutions and mixed-initiative tool design, a central theme in optimizing human–LLM collaboration is determining the appropriate level of \textit{steerability}, which \citet{Feng2024-xl} define it as the degree of control and feedback that humans have to guide an agent automated behaviors.
Research in related creative domains highlights a nuanced relationship between user control and automation. On one hand, excessive automation can undermine user sense of agency and ownership; on the other hand, simply offering maximal control does not necessarily lead to a better user experience. For example, in creative writing, \citet{Draxler2024-ii} describe the ``AI ghostwriter effect,'' where highly automated LLM-assisted writing make users feel detached from the content, a pattern similarly observed in studies \citep{Kreminski2024-pf, Joshi2025-hp}. Recent large-scale crowdsourcing studies also proved that LLM assistance could inadvertently hinder independent creative performance in everyday brainstorming, producing homogenized ideas \citep{Kumar2025-og, Qin2025-lw}.

\added{Beyond this general trend, design research in creative artwork provides a more specific insight: in human collaborative drawing with AI, \citet{Oh2018-ha} found that users reported significantly higher satisfaction only when they explicitly requested AI automation, guidance, and explanations, emphasizing the value of user-initiated actions. Similar increases in user satisfaction were observed in other LLM-assisted activities outside of creative domains.} For example, \citet{Khurana2025-xr} developed a guided copilot for built-in applications that provides actionable, learning-by-doing guidance alongside data analysis and visual design. \citet{Ma2025-fk} created an interactive LLM-based coding assistant that scaffolds and adapts to programmer personalized coding. \citet{Feng2024-xl} developed \textit{Cocoa}, a high-steerable interface enabling users to co-plan and co-execute tasks with LLM agents in a document writing environment. Across these studies, the new systems were consistently preferred over baseline chat interfaces, primarily because, as reported, \added{LLM assistance was integrated by adapting to existing user workflows that center on human participation and effort.} \deleted{they reduced the cognitive effort required to process and interpret LLM actions in domain-specific tasks that demand granular human controls.} 

\subsection{Summary and Study Focus} \added{With AI becoming more deeply integrated into scientific discovery \citep{Luo2025-bv, Liang2025-ss}, where research ideation is a central human activity reflecting scientific inquiry and reasoning (Section \ref{subsec:scientific_creativity}), and with the rise of continuous, cross-activity human-AI collaboration enabled by LLM agents (Section \ref{subsec:design_spectrum}), an important gap remains: we know little about the potential tensions between human control and AI automation in this emerging design paradigm (Section \ref{subsec:steerability_influence}).} \deleted{In the context of research ideation, measuring the influence of LLM warrants more careful consideration.} 

\added{Conceptually, we can anticipate following tensions based on insights from prior work; for example:} because the degree of controls over LLM agents reconfigures the creative labor division between humans and AI \added{across an end-to-end ideation workflow}, this might directly shape potential trade-offs between factors such as automation gains, human agency, cognitive engagement, idea novelty and feasibility \citep{Li2025-yj}. Furthermore, these varying levels affect how researchers perceive their own vs. AI contributions, raising institutional concerns about scientific integrity \citep{Bozkurt2024-ds, Editorials2023-hd}, issues of credit and ownership \citep{He2025-af}. \deleted{With the increasing use of AI in research activities that represent core human activities in research ideations, such as synthesizing literature and identifying research gaps, it becomes critical to understand how perceptions of human ownership over ideas persist across different levels of automation. Additionally, it is equally important to identify factors that motivate researchers to classify work as human-generated, AI-generated, or co-created, as these perceptions can influence their willingness to adopt and build upon AI-generated ideas.}

\added{To empirically detail these tradeoffs,} our study introduces an agentic ideation system \added{that scaffolds an end-to-end ideation workflow }with three control levels, i.e., Low, Medium, and Intensive, \added{where the availability of user-initiated interactions increases from Low to Intensive (detailed in Section \ref{subsubsec:features})}. We invited researchers from diverse disciplines to interact with the agentic system under different controls and examined how they influence the balance between LLM creativity support and researcher own effort, as well as how researchers attribute contributions and assign ownership to the final co-created ideas. 

\section{Design Process} \label{sec:design}

Given the wide design spectrum for research ideation (Section \ref{subsec:design_spectrum}), specifying appropriate design and task contexts is essential for distinguishing levels of LLM controls to assess their effects. In this section, we describe how we situate an end-to-end research ideation in the context of proposal writing (Section \ref{subsec:design-goals}), synthesize prior work and iterative design insights to inform final design choices (Section \ref{subsec:discover-define}), and develop the agentic system (Section \ref{subsec:develop-deploy}).

\subsection{Frame Design Goals} \label{subsec:design-goals}





Our research questions examine how varying control levels, reflecting different human supervision over LLM agents, influence the trade-offs between creativity support and human effort when elaborating a research idea into a concrete draft, as well as the distribution of perceived contribution and ownership between humans and AI. Notably, these questions serve as intermediate inquiries that seek to clarify how users navigate the potential tensions introduced by different degrees of automation, rather than prescribing an optimal user experience for any specific control level. This focus positions our work differently from standard user-needs-driven design. Recognizing this subtle distinction, we adopted a meta-level approach in designing our system prototypes to support the trade-off analysis. 

We grounded our Design Goals (DGs) in prior design studies and literature, and worked closely with pilot participants across multiple design stages as a \textit{research-through-design} process \citep{Zimmerman2014-cs, Zimmerman2007-gh}. Four external researchers participated in this process: three PhD students, including two studying HCI, and one in Chemistry, and one postdoctoral fellow in Physics.




\subsubsection{\textbf{DG1: Develop an agentic ideation system that supports the core activities of research proposal writing}} \label{DG:first} As highlighted by a systematic review of LLM-supported ideation by \citet{Li2025-yj}, encompassing research ideation study from both HCI \citep{Pu2025-us, Liu2024-vc} and AI communities \citep{Nigam2024-jc, Garikaparthi2025-fs}, proposal writing provides a representative academic context for examining the end-to-end, LLM-supported idea elaboration process. 

Correspondingly, we aligned our first design goal with identified user needs from \citet{Pu2025-us}, including ``help users develop early-stage ideas into concrete research briefs,'' ``explore and evaluate alternative directions,'' and ``enhance idea development by grounding LLM responses in relevant literature.'' By conducting a comparative design review from both AI and HCI work (see Section \ref{subsec:discover-define}), reviewed insights informed the formulation of three agentic roles, \textit{Ideator}, \textit{Writer}, and \textit{Evaluator}, to support the following key sub-activities: (1) generating seed ideas, (2) developing those ideas into concrete briefs grounded in literature, research goals, and study plans, and (3) evaluating the resulting briefs based on criteria tailored to researcher domains. We detail how our back-end system design (Section \ref{subsubsec:agents}) and front-end user experience (Section \ref{subsubsec:features}) meet this goal.

\subsubsection{\textbf{DG2: Design a steerable interface that supports varying control levels for trade-off analysis}} \label{DG:second} While the first design goal could be primarily driven by identified user needs, the second goal is rooted in our research inquiry to explore the tensions and trade-offs associated with different control levels.

Through close collaboration with pilot participants, we conceptualized three levels of human supervision over LLM agents, \textit{Low}, \textit{Medium}, and \textit{Intensive}, where user-initiated actions increased along the levels: the \textit{Low} level reflects minimally human supervision and highest AI automation, where the system streamlines the core user activities described in the first design goal; progressing toward the \textit{Medium} and \textit{Intensive} levels, more agency and decision-making responsibility are delegated to the human researcher. We mapped control features identified in prior mixed-initiative tool designs onto similar agent automation. This mapping yielded different fine-grained interaction capabilities that users gain at each control level, enabling the trade-off analysis (see Section \ref{subsubsec:features} for details).

\subsubsection{\textbf{DG3: Ensure system transparency through revision tracking and safe undo mechanisms}} \label{DG:third} This goal arose from both discussions within our research group and pilot testing. 

As different control levels were applied to each proposal stage, we observed that increasing automation, especially for the low control level, creates a strong user need for clear and reliable mechanisms to track changes, review revision history, and revert outputs when necessary. This design consideration also reflects user needs identified in studies of agentic systems across other activities, such as planning \citep{Feng2024-xl}, coding \citep{Ma2025-fk}, and data analysis \citep{Chen2025-ep}, where transparency and recoverability are similarly emphasized as crucial for supporting user trust and oversight. We provide these revision features available by default at all control levels, beyond those used to examine trade-offs. Section \ref{subsubsec:implementation} provides further design details to fulfill this goal.

\medskip We follow a classic double-diamond design framework to scaffold our design process \citep{Design_Council2015-wu}. This involves steps of \textit{Discover and Define}, demonstrating how we explore and finalize design specifications (Section \ref{subsec:discover-define}), and \textit{Develop and Deliver}, representing how we prototype and deploy the design concept (Section \ref{subsec:develop-deploy}). 
While the double-diamond design is presented as a streamlined procedure, the actual development 
is typically iterative and participatory in nature \citep{Laurel2003-ja}, involving internal and external evaluations. Accordingly, we iteratively developed and refined our design concept at each stage \citep{Zimmerman2007-gh, Zimmerman2014-cs}. Besides detailing the design process, we illustrate how design goals were met.

\begin{table*}[t]
\centering
\resizebox{\linewidth}{!}{%
\begin{tabular}{@{}>{\RaggedRight}p{3.2cm} >{\RaggedRight}p{2cm} >{\RaggedRight\arraybackslash}p{10cm}@{}}
\toprule
\textbf{Agentic System} & \textbf{Components} & \textbf{Main Tasks and Corresponding Tool Use} \\
\midrule
\textbf{SciAgents} \citep{Ghafarollahi2024-tl} 
& Ontologist & Connect research concepts based on a knowledge graph and articulate concept relationships \\
& Scientist 1 \& 2 & Collaboratively develop and iteratively improve the idea based on prior concepts into a full research proposal \\
& Critics & Summarize the proposal, highlight strengths and weaknesses, and suggest improvements \\
\midrule
\textbf{The AI Scientist} \citep{Lu2024-jg} 
& Idea Generation & Generate seed ideas, assess novelty against the literature, and select the most promising idea based on interestingness, novelty, and feasibility \\
& Experiment Iteration & Expand the idea with experimental plans, develop a codebase to test the hypothesis, and visualize results \\
& Paper Write-up & Generate and refine a manuscript based on experimental results and incorporate references \\
\midrule
\textbf{Agent Laboratory} \citep{Schmidgall2025-ff} 
& Literature Review & Collect relevant papers for a given idea and summarize them into a curated review \\
& Experimentation & Design an actionable research plan grounded in the literature and aligned with the research goal; identify relevant datasets or codebases; and execute the plan \\
& Report Writing & Synthesize findings into a report, refine through iterative edits, add references, and review according to conference guidelines \\
\midrule
\textbf{AI Co-Scientist} \citep{Gottweis2025-hh} 
& Generation & Explore literature and simulate debate to generate hypotheses and proposals aligned with research goals \\
& Reflection & Assess quality, novelty, and correctness of hypotheses and provide improved explanations via literature search \\
& Ranking & Evaluate and prioritize proposals using pairwise comparisons and simulated debates \\
& Proximity & Compute proximity graphs to cluster similar ideas \\
& Evolution & Refine top-ranked hypotheses by synthesizing ideas, drawing analogies, and simplifying complex concepts \\
& Meta-review & Iteratively improve the quality and relevance of hypotheses and reviews \\
\bottomrule
\end{tabular}
}
\vspace{0.2cm}
\caption{Agent components and main tasks in four prior end-to-end scientific discovery systems. In comparison, SciAgents \citep{Ghafarollahi2024-tl} and Agent Laboratory \citep{Schmidgall2025-ff} follow a unified, single-path refinement process, The AI Scientist \citep{Lu2024-jg} and AI Co-Scientist \citep{Gottweis2025-hh} adopt multi-turn screening and ranking workflows that generate and evaluate multiple research ideas before selection.}
\Description{Agent components and main tasks in four prior end-to-end scientific discovery systems. In comparison, SciAgents \citep{Ghafarollahi2024-tl} and Agent Laboratory \citep{Schmidgall2025-ff} follow a unified, single-path refinement process, The AI Scientist \citep{Lu2024-jg} and AI Co-Scientist \citep{Gottweis2025-hh} adopt multi-turn screening and ranking workflows that generate and evaluate multiple research ideas before selection.}
\label{tab:agent-components}
\end{table*}

\subsection{Discover and Define the Agentic Workflow} \label{subsec:discover-define}

To identify how to create an agentic ideation system that effectively supports proposal writing (DG1 \ref{DG:first}) and different control levels to explore trade-offs (DG2 \ref{DG:second}), we draw insights from the design spectrum, ranging from fully automated to mixed-initiative approaches. Below, we summarize the prior design knowledge that informs our system development.

\subsubsection{\textbf{Identifying agentic components to fulfill the back-end requirements of DG1}} \label{subsubsec:agents} In this section, we outline main agentic components (e.g., roles, tools, and memory use) that fulfill the backend system requirements for DG1 (\ref{DG:first}). Drawing on insights of four prominent end-to-end scientific discovery system (summarized in Table \ref{tab:agent-components}), our system adopts their core agentic components, such as literature exploration and synthesis, hypothesis generation, and study planning, to support the specific agent activities in research proposal writing.


Key differences exist among prior agentic workflows and outputs. For example, \textit{SciAgents} \citep{Ghafarollahi2024-tl} and \textit{Agent Laboratory} \citep{Schmidgall2025-ff} adopt a unified, unidirectional workflow to iteratively refine a single research idea from conception to execution and writing. In contrast, \textit{the AI Scientist} \citep{Lu2024-jg} implements a multi-turn, screening-based workflow that generates, evaluates, and ranks multiple seed ideas before selecting one for further development. \textit{AI Co-Scientist} \citep{Gottweis2025-hh} extends this by supporting iterative ranking and merging of different seed ideas, ultimately producing a set of top-ranked research proposals. 

\textbf{Agents.} Our study seeks to better understand human involvement in research ideation, situating this inquiry within a realistic proposal writing scenario. Instead of replicating a large-scale idea filtering workflow used in \textit{The AI Scientist} and \textit{AI Co-Scientist}, we adopt a streamlined workflow that is less complex but still representative of real-world practice. Following the design of \textit{SciAgents} and \textit{Agent Laboratory}, our approach retains standard agent roles: 
\begin{itemize}
    \item \textbf{Ideator}: Generate seed ideas with a brief abstract based on keywords provided by researchers
    \item \textbf{Writer}: Expand a seed idea with different proposal sections, including literature synthesis, research goals, and study plan
    \item \textbf{Evaluator}: Provide critiques and suggestions for improvements to the existing proposal
\end{itemize}

We then integrate these three roles into a fully automated workflow. Specifically, the \textit{Ideator} and \textit{Writer} can automatically incorporate critiques and suggestions provided by the \textit{Evaluator} to generate an improved seed idea and revise the proposal.

\textbf{Tools.} We primarily employ academic web search using platforms such as Semantic Scholar and arXiv\footnote{Semantic Scholar datasets: \url{https://api.semanticscholar.org/api-docs/datasets}; arXiv API: \url{https://info.arxiv.org/help/api/index.html}}. Additionally, we incorporate open web search to assist with query expansion, improving the relevance and breadth of retrieved content.

\textbf{Memory.} We adopt a dual-memory architecture \citep{Weng2023-lr}. The short-term memory maintains relevant academic context and proposal drafts within a single iteration session, retaining the current version along with the immediately preceding one. Long-term memory is a processed knowledge base derived from broader web search results, enabling the system to access over accumulated external information across sessions.

Further technical details are presented in Section \ref{subsec:develop-deploy}.

\begin{table*}[t]
\centering
\resizebox{\linewidth}{!}{%
\begin{tabular}{>{\RaggedRight}p{3.5cm} >{\RaggedRight}p{6cm} >{\RaggedRight\arraybackslash}p{6cm}@{}}
\toprule
\textbf{Mixed-initiative Tools} & \textbf{User Activities and Controls} & \textbf{Corresponding LLM Features} \\
\midrule
\textbf{CoQuest} \citep{Liu2024-he} & 
Generate either breadth-first or depth-first research questions;  \newline
Provide feedback to AI for refining research questions; \newline 
Explore paper graph & 
Retrieve relevant papers based on user keywords and generate research questions; \newline 
Use knowledge graph to show connected papers relevant to the keywords \\
\midrule
\textbf{PersonaFlow} \citep{Liu2024-vc} & 
Generate research ideas by customizing expert personas or uploading important literature & 
Prompt LLM as different expert personas to ideate research questions \\
\midrule
\textbf{Scideator} \citep{Radensky2024-gq} & 
Provide research topics, attach relevant papers, and select criteria facets;  \newline
Evaluate and select ideas & 
Retrieve new papers based on facet distance and generate ideas;  \newline
Use GPT-based ranker to prioritize and organize papers under different ideas and assess idea novelty \\
\midrule
\textbf{IdeaSynth} \citep{Pu2025-us} & 
Search and add papers for collection;  \newline
Specify seed idea facets for solutions, methods, evaluations, or impacts;  \newline
Select multiple facet nodes and generate a research brief & 
Generate research literature summary and analysis based on user-selected papers;  \newline
Generate corresponding results for different facets of a seed idea;  \newline
Compile user-selected facets to generate a research brief \\
\midrule
\textbf{Cocoa} \citep{Feng2024-xl} & 
Initiate a plan request for agents, edit plans, and assign steps for agents to execute;  \newline
Edit agent outputs & 
Generate a plan based on user prompts, search the web to execute the plan;  \newline
Provide replan options based on user edits \\
\midrule
\textbf{IRIS} \citep{Garikaparthi2025-fs} & 
Brainstorm research ideas in a chat interface and search/attach relevant papers to assist idea generation & 
Retrieve relevant papers and automatically generate a research brief with evaluated scores across different criteria \\
\bottomrule
\end{tabular}
}
\vspace{0.2em}
\caption{User controls and corresponding LLM features across prior mixed-initiative research ideation tools. Tools vary in both output and interface design, ranging from question- or hypothesis-focused brainstorming (e.g., \textit{CoQuest} \citep{Liu2024-he}, \textit{Scideator} \citep{Radensky2024-gq}) to comprehensive research briefs with customizable sections, and from flexible ideation boards (e.g., \textit{PersonaFlow} \citep{Liu2024-vc}, \textit{IdeaSynth} \citep{Pu2025-us}) to structured, chat-based, or notebook-style environments (e.g., \textit{IRIS} \citep{Garikaparthi2025-fs}, \textit{Cocoa} \citep{Feng2024-xl}).}
\Description{User controls and corresponding LLM features across prior mixed-initiative research ideation tools. Tools vary in both output and interface design, ranging from question- or hypothesis-focused brainstorming (such as \textit{CoQuest} \citep{Liu2024-he}, \textit{Scideator} \citep{Radensky2024-gq}) to comprehensive research briefs with customizable sections, and from flexible ideation boards (\textit{PersonaFlow} \citep{Liu2024-vc}, \textit{IdeaSynth} \citep{Pu2025-us}) to structured, chat-based, or notebook-style environments (\textit{IRIS} \citep{Garikaparthi2025-fs}, \textit{Cocoa} \citep{Feng2024-xl})}
\label{tab:user-controls-llm-features}
\end{table*}

\subsubsection{\textbf{Defining controls to support the front-end user experience of DG1 and separating supervision levels for DG2 trade-off analysis.}} \label{subsubsec:features}

\added{Drawing on use cases from prior mixed-initiative research ideation systems, we outline discrete user activities with controls. This enables the front-end user experience of a coherent, end-to-end idea elaboration process for DG1 (\ref{DG:first}). Additionally, we conceptualized three supervision levels (Low, Medium, and Intensive) and mapped use features across the proposal stages to these levels, fulfilling the DG2 (\ref{DG:second}).}

Table~\ref{tab:user-controls-llm-features} summarizes the main user activities with controls and corresponding LLM features in recent mixed-initiative tools developed for research ideation. Several key differences exist. First, they produce different types of ideation outputs, reflecting distinct study goals. For example, \textit{CoQuest} \citep{Liu2024-he} and \textit{Scideator} \citep{Radensky2024-gq} focus on generating research questions or hypotheses, supporting early-stage divergent brainstorming. In contrast, other tools generate a more comprehensive research brief assembling different sections, including literature synthesis, proposed methodologies, and study plans. These tools allow greater user input, such as directly editing outputs or customizing prompts for each section.

Additionally, these tools differ in interface design. \textit{CoQuest} \citep{Liu2024-he}, \textit{PersonaFlow} \citep{Liu2024-vc}, and \textit{IdeaSynth} \citep{Pu2025-us} adopt a format of ideation boards that enable flexible manipulation of nodes. In contrast, \textit{Scideator} \citep{Radensky2024-gq} employs a highly structured interface centered on its facet-filtering mechanism for idea evaluation. \textit{IRIS} \citep{Garikaparthi2025-fs} defaults to a chat-based interface, while \textit{Cocoa} \citep{Feng2024-xl} adopts a notebook writing environment with greater flexibility for customization and editing.

\textbf{Main User Activities for DG1.}
Building on prior work, we synthesize four main user activities powered by the agentic system, reflecting key tasks researchers typically perform when developing a research proposal.
\begin{enumerate}
    \item \textbf{Brainstorming Seed Ideas}: Researchers initiate the process by specifying research topics, while agents assist in formulating novel seed ideas or hypotheses by comparing with existing work. Researchers could edit and select the most promising ideas to pursue.
    \item \textbf{Exploring and Synthesizing Literature}: Researchers identify relevant literature through web search, and agents assist by synthesizing key findings and articulating how they support or motivate the selected seed idea.
    \item \textbf{Revising Research Goals and Study Plans}: Agents generate initial drafts of research goals and study plans, and researchers steer the process through prompting or directly revise the outputs.
    \item \textbf{Co-critiquing the Proposal}: Researchers and agents collaboratively critique and iteratively revise the proposal.
\end{enumerate}

\begin{table*}[t]
\centering
\resizebox{\linewidth}{!}{%
\begin{tabular}{@{}>{\RaggedRight}p{2.3cm}>{\RaggedRight}p{5cm}>{\RaggedRight}p{5cm}>{\RaggedRight\arraybackslash}p{5cm}@{}}
\toprule
\textbf{User Activities} & \textbf{Intensive} & \textbf{Medium} & \textbf{Low} \\
\hline
Brainstorming seed ideas & 
1. Provide keywords to generate and select seed ideas \linebreak
2. Customize seed ideas &
1. Provide keywords to generate and select seed ideas \linebreak
2. Customize seed ideas &
1. Provide keywords to generate and select seed ideas \linebreak
2. Customize seed ideas \\
\hline
Exploring and synthesizing \linebreak literature &
1. Search and select literature \linebreak
2. Prompt agents for sectional changes \linebreak
3. Highlight texts for LLM edits \linebreak
4. Edit specific content
&
1. Search and select literature \linebreak
2. Prompt agents for sectional changes
&
\multirow[c]{3}{4.7cm}{
  \parbox[c][\dimexpr8\baselineskip][c]{4.7cm}{%
    \textit{No user involvement; fully automated by agents for literature selection, research goals, study plan generation, and critique/improvement selection.}%
  }%
}
\\
\cline{1-3}
Revising research goals and study plan & 
1. Search and select literature \linebreak
2. Prompt agents for sectional changes \linebreak
3. Highlight texts for LLM edits \linebreak
4. Edit specific content
&
1. Search and select literature \linebreak
2. Prompt agents for sectional changes
&
\\
\cline{1-3}
Co-critiquing the proposal & 
1. Customize evaluation criteria \linebreak
2. Provide feedback for agent critiques \linebreak
3. Select and customize agent-suggested improvements
&
1. Customize evaluation criteria \linebreak
2. Select agent-suggested improvements
&
\\
\bottomrule
\end{tabular}
}
\vspace{0.2em}
\caption{Mapping user features across stages of proposal writing for three control levels: Intensive, Medium, and Low.}
\Description{Mapping user features across stages of proposal writing for three control levels: Intensive, Medium, and Low.}
\label{tab:feature mapping}
\end{table*}

\textbf{User Controls and Supervision Levels for DG2.} Building on the idea of steerability \citep{Feng2024-xl}, separating user controls for various forms of human oversight should mirror how much flexibility users have to intervene in LLM agents throughout the above activities.

By co-designing with pilot participants, we arrived at the following three control levels, reflecting progressively greater human supervision over agents:
\begin{description}
    \item[Low] This level should reflect a minimally supervised agentic workflow, where humans initiate the process, after which idea elaboration proceeds entirely by agents. The system autonomously iterates on the proposal until the researcher is satisfied, allowing minimal oversight with only a few degrees of freedom for intervention.
    \item[Medium] This level should reflect a partially supervised agentic workflow, where humans can steer high-level direction and adjust proposal focus, while agents continue to elaborate idea details. In this level, we assume researchers accept the unpredictability of LLM outputs, since they cannot directly refine the generated text except as a manual post-processing step outside of the system.
    \item[Intensive] This level should reflect a fully supervised agentic workflow, where humans maintain full control over both high-level direction and fine-grained content development. Agents assist with iterative refinement. With maximal human control, this allows for iterative and highly collaborative proposal development between humans and AI.
\end{description}


\added{Guided by the conceptualization described above, we mapped user features across proposal stages and categorized them according to supervision levels, as summarized in Table \ref{tab:feature mapping}. Across user activities, all control levels allow researchers to specify keywords, then generate and refine seed ideas at the ideation stage (a function commonly shared across prior agentic systems from full automation to mixed-initiative tools). Differences exist during the subsequent idea elaboration process: \textit{Intensive} control gives researchers full agency over literature search, content editing, and evaluation; \textit{Medium} control provides selective oversight at the sectional level without direct content-level edits and evaluation updates; and \textit{Low} control relies primarily on automation, with the agents performing all tasks to elaborate a seed research idea. This structured separation satisfies DG2 (\ref{DG:second}) to enable a trade-off analysis.}

\subsection{Develop and Deploy the Agentic System} \label{subsec:develop-deploy}

Table \ref{tab:feature mapping} serves as a foundation for us to implement a functional software prototype, as well as enabling feature separation of different control levels. During this process, we conducted pilot user testing with external researchers to design the user interface (UI), identify potential usability issues, and revise study protocols for formal user studies. Additionally, feedback from pilot participants informed the development of features to satisfy DG3 (\ref{DG:third}), including revision tracking and a safe undo mechanism to ensure system transparency. In the remainder of this section, we describe the UI (Section \ref{subsubsec:UI}) and the technical implementation (Section \ref{subsubsec:implementation}).

\subsubsection{User interface} \label{subsubsec:UI}

\begin{figure*}[ht]
    \centering
    \includegraphics[width=\linewidth]{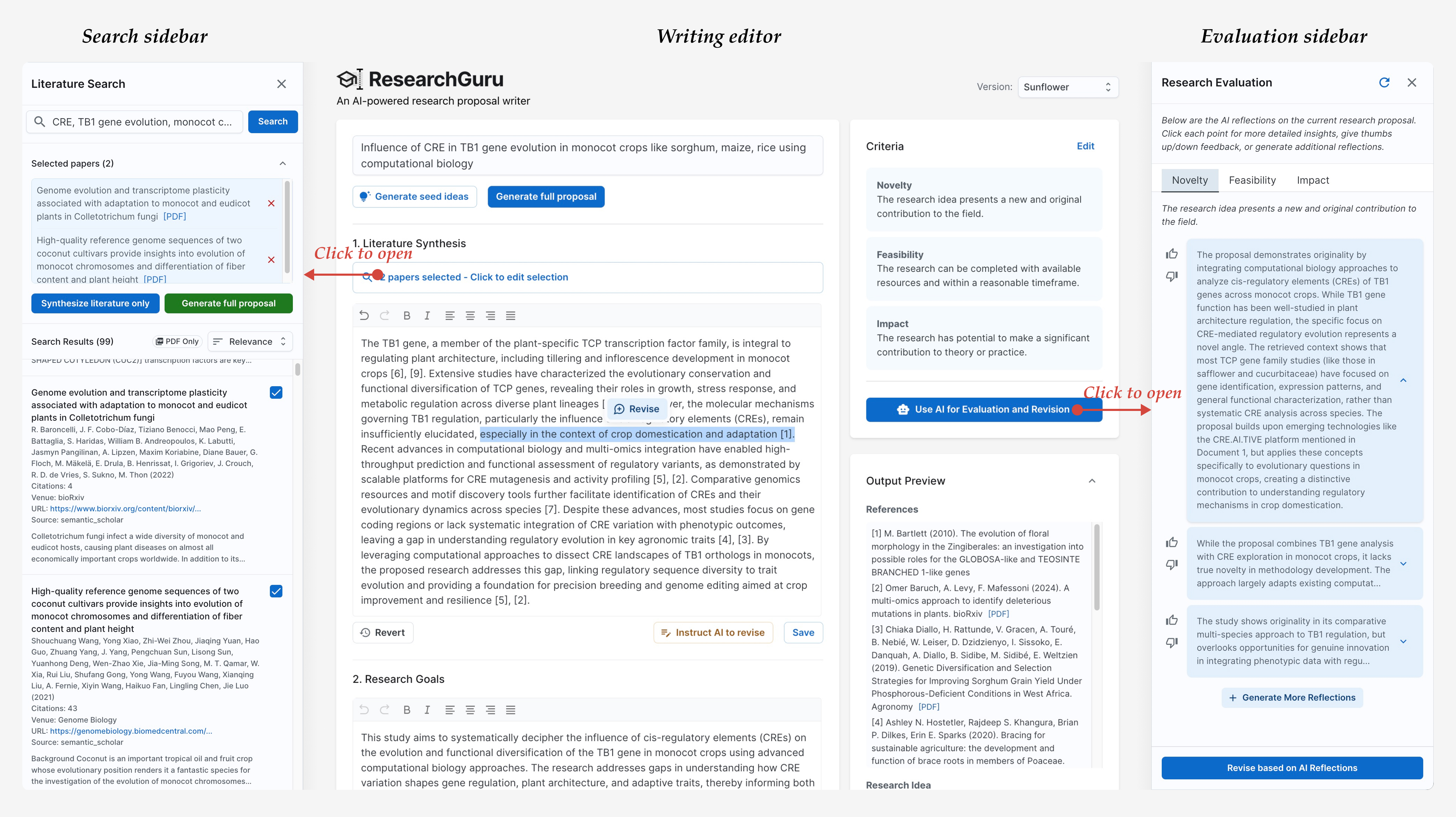}
    \caption{Screenshots of the \textit{ResearchGuru} system dashboard. Users primarily work in the \textit{Writing Editor}, where they can both initiate functions and edit their proposals. The \textit{Search Sidebar} and \textit{Evaluation Sidebar} can be opened on the left and right sides of the dashboard, respectively.}
    \label{fig:main_UI}
    \Description{main UI}
\end{figure*}

The system UI uses a notebook-style text editor (Figure \ref{fig:main_UI}), inspired by GPTCanvas, LangChain Open-Canvas\footnote{GPTCanvas: \url{https://openai.com/index/introducing-canvas/}; Open-Canvas: \url{https://github.com/langchain-ai/open-canvas}}, and \textit{Cocoa} \citep{Feng2024-xl} to support proposal writing. Like \textit{Cocoa} and unlike GPTCanvas and Open-Canvas, we excluded a chat feature because our focus is on sustained co-writing with LLM agents rather than conversational interactions. Our system supports section-specific and in-line text prompting, and retains core features such as local editing, paper search, and interactive proposal evaluation. 

\begin{figure*}[ht]
    \centering
    \includegraphics[width=\linewidth]{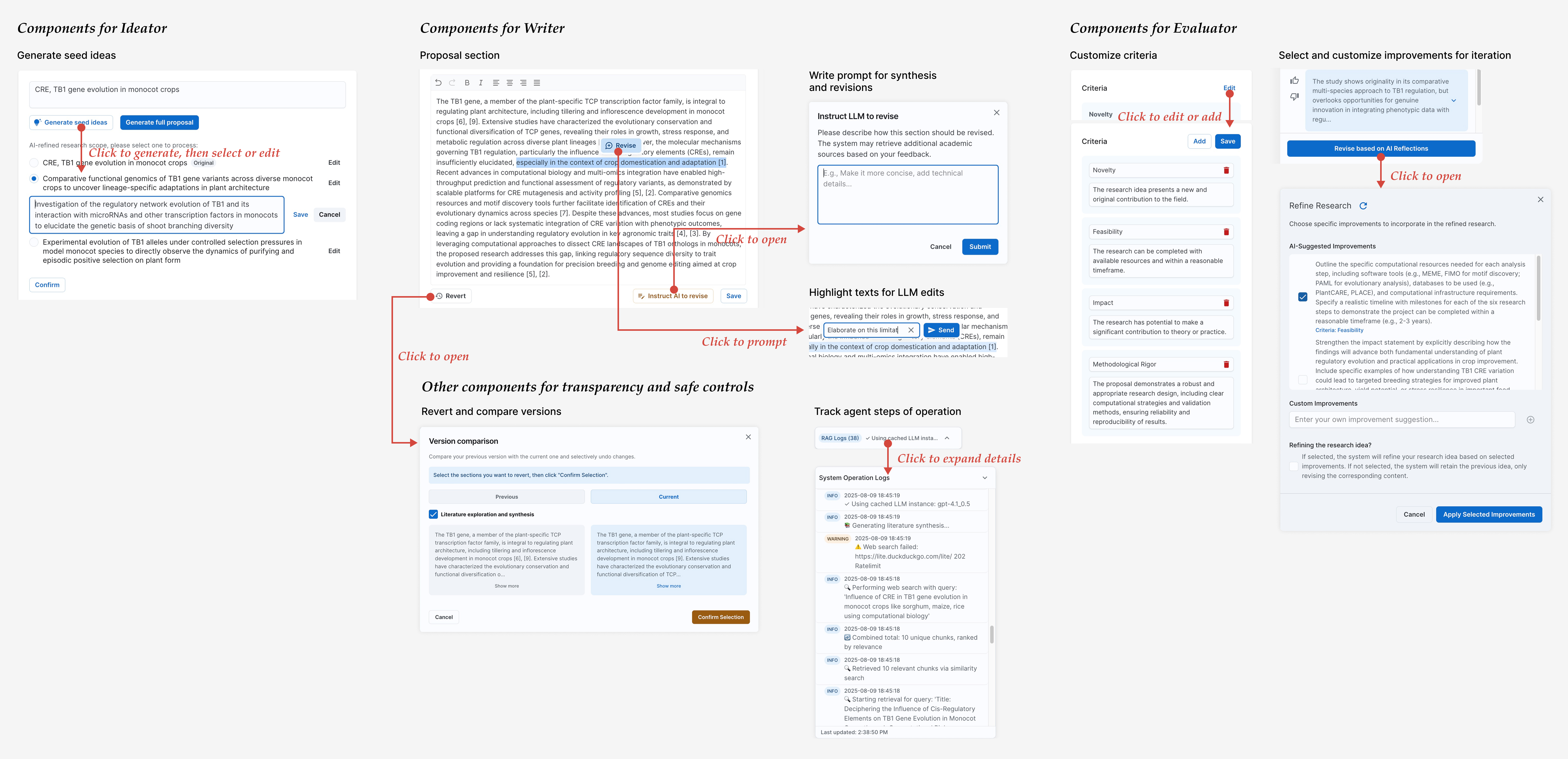}
    \caption{Screenshots of different agentic functions for the \textit{ResearchGuru} system dashboard. Users can select and edit agent suggested seed ideas; write prompt for synthesis and revisions, or highlight texts for LLM edits in each proposal section; customize criteria, select and customize agent suggested improvements; and revert versions for each section and review agent steps of operations.}
    \label{fig:components_UI}
    \Description{Details of UI components}
\end{figure*}

Based on the \textit{User Controls} (Section \ref{subsubsec:features}), we design corresponding UI components and pair them with different \textit{Agents} (Section \ref{subsubsec:agents}) (Figure \ref{fig:components_UI}). Focusing on the \textbf{Intensive} steering condition for the moment, we present an idealized use scenario below describing how researchers might interact with this UI:

\begin{description}
    \item[Phase one:]\textbf{Formulate seed ideas for proposal.} Researchers begin by exploring potential research directions based on keywords. The \textit{Ideator} reviews relevant existing articles and suggests possible seed ideas. Researchers can then edit or select a specific idea or add extra information, such as datasets or methodologies. Once sufficient details are included, they click ``Generate\_Full\_Proposal'' and the \textit{Writer} starts drafting each proposal section.
    \item[Phase two:]\textbf{Revise and edit proposal.} The \textit{Writer} generates the initial draft by integrating relevant papers into sections: literature synthesis, research goals, and study plan. For each, users can prompt the \textit{Writer} how to compose, highlight text to revise, or make direct edits themselves. They can also search and select specific articles to include.
    \item[Phase three:]\textbf{Evaluate proposal.} Researchers customize the criteria to evaluate the proposal. The \textit{Evaluator} then generates critiques by comparing the research proposal with the literature. Researchers can request additional critiques under each criterion and use thumbs up/down to express their agreement or disagreement. Given this, the \textit{Evaluator} provides suggestions to improve the proposal. Researchers can select or customize the suggestions. By clicking ``Apply\_Selected\_Improvements,'' the \textit{Writer} then initiates the revision process based on suggestions.
\end{description}

The three phases above apply only to the Intensive condition, which allows full control. For the Medium and Low conditions, we streamline steps and reduce user controls as follows:
\begin{description}
    \item[Medium] This level supports only prompt-based revisions and fixed suggestions. In Phase two, researchers prompt the \textit{Writer} to revise each section, but cannot make direct edits or use text highlighting to trigger LLM-based revision. In Phase three, they cannot provide feedback to the \textit{Evaluator} and can only choose predefined agent suggestions. 
    \item[Low] This level supports only basic prompting. Researchers write prompts to the \textit{Ideator}, after which the writing and evaluation proceed automatically, without any steering controls over \textit{Writer} and \textit{Evaluator}. They can only accept the existing proposal and the agent-suggested revision.
\end{description}

\added{\textbf{Revision Tracking and Undo Features for DG3.}} Based on the different user experience outline above, we created two features, available in all control levels to improve transparency and version safety for DG3 (Section \ref{DG:third}). A revert UI modal (i.e., a pop-up window) lets researchers review and revert versions for each proposal section, while a floating modal (i.e., a smaller, floating, expandable window) allows them to review every agent step, including both successful and failed attempts.

\subsubsection{Technical implementation} \label{subsubsec:implementation}
We define prompts for different agents and integrate them into a streamlined workflow. Each agent uses an enhanced retrieval-augmented generation (RAG) pipeline \citep{Lewis2020-yp}, leveraging specialized systems to tailor its behavior. Below, we detail each technical component.

\begin{figure}[ht]
    \centering
    \includegraphics[width=0.8\linewidth]{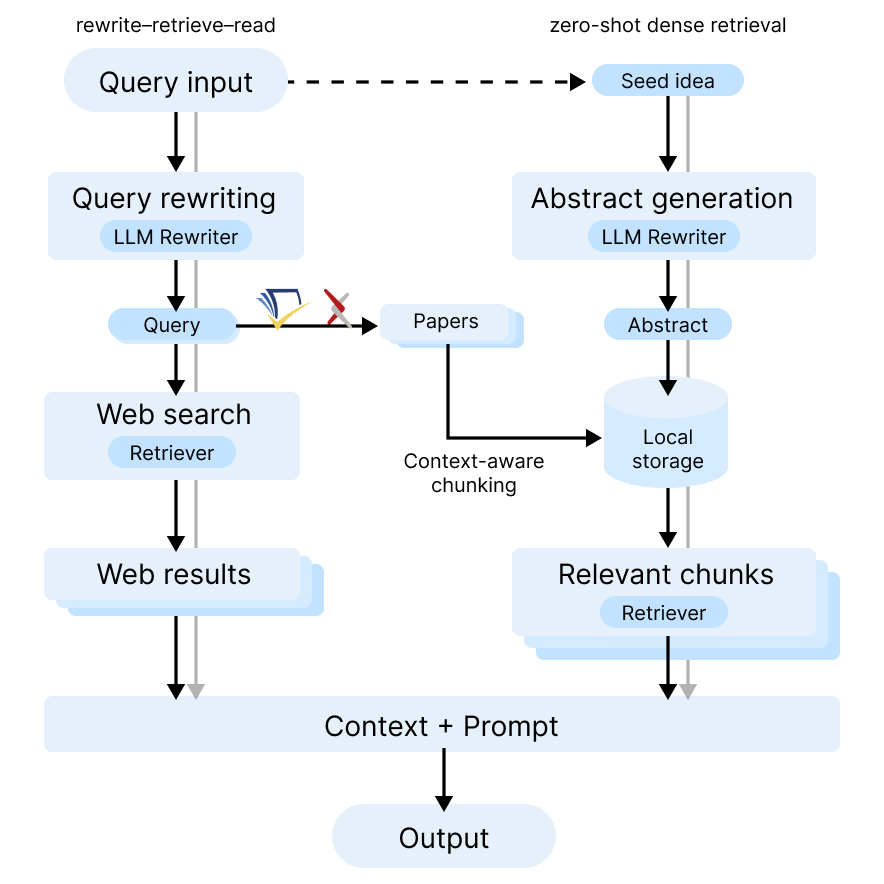}
    \caption{The RAG system integrates a rewrite-retrieve-read framework \citep{Ma2023-yk} with zero-shot dense retrieval \citep{Gao2023-eb}.}
    \label{fig:RAG_architecture}
    \Description{An enhanced rag system integrates a rewrite-retrieve-read framework with zero-shot dense retrieval}
\end{figure}

\textbf{Enhanced RAG.} Following a standard RAG workflow\footnote{Langchain RAG tutorial: \url{https://python.langchain.com/docs/tutorials/rag/}} (indexing and retrieving content to support generation based on a user query), our system adopts a rewrite-retrieve-read framework, proposed by \citet{Ma2023-yk} (see Figure \ref{fig:RAG_architecture}). Specifically, the LLM first rewrites the query, then uses a web search engine to gather curated snippets of related concepts from the web. These snippets are later incorporated into the prompt to enrich the generation process for factual grounding. The written query is also used to search papers from Semantic Scholar and arXiv. We index and embed relevant papers in a local database as long-term memory.

During the embedding, we implement \textit{context-aware chunking}\footnote{Chunking strategies for LLM applications: \url{https://www.pinecone.io/learn/chunking-strategies/}}. Academic papers typically follow a standardized structure, such as Introduction, Related Work, Methods, Results, and Discussion, that closely corresponds to our research proposal sections. Leveraging this alignment, we segment each paper into chunks based on its original sections. During retrieval, chunks that correspond to specific proposal sections (e.g., ``Methods'' in the source paper mapped to ``Study Plan'' in the proposal) are prioritized and ranked higher.

We also employ \textit{zero-shot dense retrieval} \cite{Gao2023-eb} to enhance retrieval relevance for research content. This process starts by prompting the LLM to generate a hypothetical abstract based on a user-selected seed idea. 
This abstract then serves and saves as an enhanced query associated with this seed idea to retrieve relevant content from the database.
The top-\textit{k} papers are selected and passed to the LLM along with task-specific prompts for generation.

\begin{figure}[ht]
    \centering
    \includegraphics[width=0.8\linewidth]{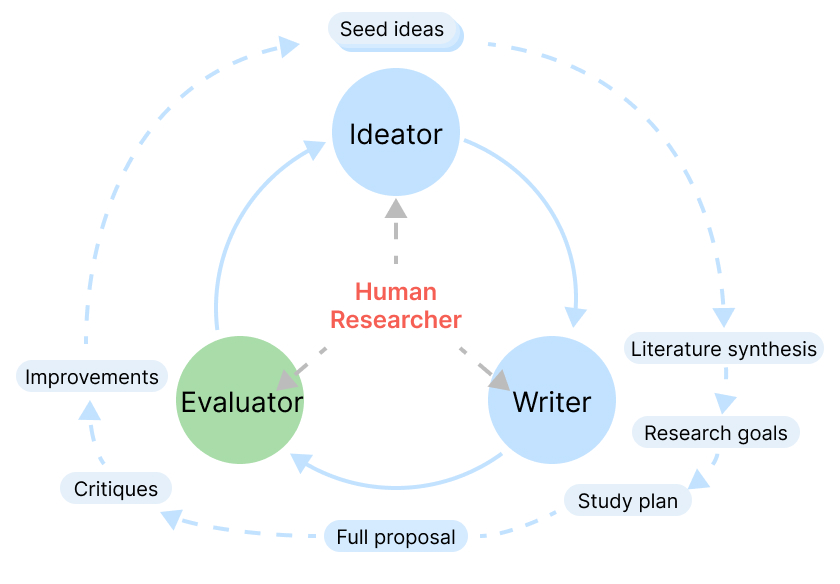}
    \caption{The agentic writing workflow involves multiple LLM calls, each generating distinct outputs through an internal chain-of-thought process. In our development, we employ two different LLMs: the Ideator and Writer roles share the same LLM, while the Evaluator utilizes a different LLM. The system allows human researchers to stop or intervene each agent with different controls.}
    \label{fig:Agentic_workflow}
    \Description{The agentic writing workflow}
\end{figure}

\textbf{Agentic Workflow.} 
As illustrated in Figure \ref{fig:Agentic_workflow}, we designed an agentic writing workflow that follows predefined paths to orchestrate different LLM calls and systems. Instead of creating a fully autonomous agent, we adopt the \textit{LLM workflow} as a simpler architecture\footnote{Building agents: \url{https://www.anthropic.com/engineering/building-effective-agents}}. In a proposal-writing context, most subtasks, such as generating ideas, expanding outlines, and reviewing drafts, have clear inputs, outputs, and evaluation criteria, making them easier to orchestrate through a structured workflow.

This workflow consists of: a) the \textit{Ideator} first generates multiple seed ideas; b) the \textit{Writer} then expands a researcher-selected idea into a full proposal; c) the \textit{Evaluator} reviews the proposal, providing critiques and suggestions for improvement; d) finally, the \textit{Ideator} and \textit{Writer} refine the idea and proposal based on the \textit{Evaluator}'s feedback. We adapt the enhanced RAG (Figure \ref{fig:RAG_architecture}) to each agentic role, as their complexity and technical requirements differ. We summarize these distinctions in Appendix \ref{appendix:prompt details} with prompt details.


\textbf{System Deployment.}
We deploy the RAG components and the agentic workflow in Python using LangChain, with Flask serving as the web framework to connect backend services to the front-end interface. The steerable interface is built using React with MUI as the UI component library\footnote{LangChain: \url{https://langchain.com/}; Flask: \url{https://flask.palletsprojects.com/}; Material UI: \url{https://mui.com/}}. To tailor the system into three control levels, we design feature variation to switch between different functionalities and streamline agent functions.

For the final system used in the formal user study, we employed two LLM services to power three agentic roles. GPT-4.1 was used for the \textit{Ideator} and \textit{Writer}, while the \textit{Evaluator} was deployed using Claude Sonnet 3.7. The primary reason for using two different LLMs was to avoid {\em self-preference bias} \cite{panickssery2024llm}--- a model preferring its own generations ---for more objective evaluation during proposal iteration. We document the system evaluation in Section \ref{sec:system-evaluation}, detailing how we assess the performance of RAG and agentic workflow for proposal writing, compared to off-the-shelf LLM services. Other services or libraries used in building our system include: Chroma for indexing and embedding, DuckDuckGo for open web search, Semantic Scholar and arXiv as academic databases, and the Lexical editor for writing\footnote{GPT-4.1: \url{https://openai.com/index/gpt-4-1/}; Claude Sonnet 3.7: \url{https://www.anthropic.com/news/claude-3-7-sonnet}; Chroma: \url{https://www.trychroma.com}; DuckDuckGo: \url{https://duckduckgo.com}; Semantic Scholar: \url{https://www.semanticscholar.org}; arXiv: \url{https://arxiv.org}; Lexical editor: \url{https://lexical.dev}.}.

\section{System Evaluation and Pilot Study} \label{sec:system-evaluation}


\subsection{System Evaluation}

We set up three default evaluation criteria: novelty, feasibility, and impact, adopted from \citet{Pu2025-us}. Novelty reflects whether ``the research idea presents a new and original contribution to the field.'' Feasibility assesses whether ``the research can be completed with available resources and within a reasonable timeframe.'' Impact considers ``the research’s potential to make a significant contribution to its domain.'' Recognizing that evaluation standards may vary across disciplines, the system also allows participants to customize these default criteria or introduce new ones.

We conducted two evaluations to assess the quality and performance of our backend system, which serves as the fundamental agent capability powering the fully automated idea elaboration process. The first examined whether integrating RAG improved proposal generation quality compared with the non-RAG version, using both human expert and LLM-as-judge assessments based on the three default criteria. Two pilot participants provided domain-specific research keywords they knew well, allowing them to more effectively assess the quality of generated content. This resulted in ten seed ideas that the agentic system expanded using \textit{zero-dense retrieval}, both with and without RAG integration (see Section \ref{subsubsec:implementation} and Appendix \ref{appendix:prompt details} for specific technical details). To ensure consistency in idea expansion for system comparison, the first abstract generated from each seed idea was used across both systems to produce the concrete briefs. Using the default idea evaluation criteria, LLM judgments favored the RAG condition in most cases but humans were uncertain in several instances because the proposals were too general. Key issues identified included over-integration of literature, reducing cohesion, and proposals that were too broad for domain-specific challenges. We addressed these by refining prompts to encourage critical use of context and increasing flexibility in citation handling.

Second, we evaluated the system against off-the-shelf agentic tools with three pilot participants from chemistry, physics, and child development. They compared proposals generated by our system, Deep Research, ChatGPT, and Gemini\footnote{Deep Research (\url{https://openai.com/index/introducing-deep-research/}), ChatGPT (\url{https://chatgpt.com/}), and Gemini (\url{https://gemini.google.com/app}), using current model versions as of early June 2025  (See Appendix \ref{appendix:system eval} for specific version).}. Rankings showed that our system performed comparably, though some cohesion and clarity issues persisted. Overall, the evaluations demonstrated that RAG integration improves content specificity, our system remains competitive with existing tools. Future iterations that refine context handling and strengthen the cohesion of proposals would help improve output quality across research domains. We provide more detailed evaluation results in Appendix \ref{appendix:system eval}.

\subsection{Pilot User Testing} \label{subsec:pilot}
We refined the study tasks and protocols through several rounds of testing with pilot participants. A between-subjects design was finally adopted to ensure clean comparisons across experimental conditions. This helped us avoid the carryover and learning effects observed in pilot within-subjects design.

Several issues were reported via a within-subject approach. First, research proposal writing was a cognitively demanding task. Pilot participants reported that both thinking research questions and editing proposals required substantial time and effort. Such cognitive load introduced carryover effects in the within-subjects procedure, as three experimental conditions substantially increased the risk of participant fatigue, which is also a critical concern for creative tasks. Second, pilot participants reported that they became more productive only after familiarizing themselves with the system and organizing their ideas. This demonstrated a learning effect; they tended to prefer ideas generated in later stages, confounding comparisons across control levels. We therefore transitioned to a between-subjects experimental design and observed high improvements in both user experience and outcomes.

The three-conditions, within-subjects pilot tests resulted in more than two hours of work on average to generate one proposal under each condition (with a 5 minute break provided midway). To further prevent idea burnout and participant fatigue, we intentionally set and carefully managed the study time limits within the between-subjects procedure. Participants were asked to focus on the core components of a research proposal, e.g., how the idea relates to prior work, the main research questions, and the overall research plan, rather than on detailed writing or stylistic polishing. By narrowing the focus in this way, we were able to shorten the overall task time to 90 minutes. 

\section{User Study} \label{sec:user-study}

\subsection{Study Procedure} \label{subsec:procedure}

The user study protocol consists of four steps in which participants:
\begin{enumerate}
    \item Tool Familiarization: Familiarized themselves with the tool by reviewing either a text-based or video tutorial and used one prior research idea to briefly explore the tool and get hands-on experience
    \item Idea Elaboration Task: Created new proposals aligned with their current research interests using one control level
    \item Post-survey: Completed a post-task survey to assess perceived creativity support and human effort  
    \item Post-interview: Participated in a follow-up interview to reflect on their use experience    
\end{enumerate}

During the user study, participants were randomly assigned to Low, Medium, or Intensive control levels while using the system to develop research ideas across all proposal stages. They were asked to generate new research ideas within their areas of expertise. In the familiarization phase, participants were allowed to ask any questions regarding tool use and research idea focus. They then completed the idea elaboration task individually and could choose to end the session early or continue working until they were satisfied with the draft.
The entire study lasted approximately 90 minutes, consisting of about 20 minutes for the familiarization phase, 35-40 minutes for the proposal generation task, 5 minutes for the post-task survey, and 20-30 minutes for the interview.
Additional study protocol details in Appendix \ref{appendix:study protocol}. We also present a representative case illustrating tool usage in Appendix \ref{appendix:representative case}.

\subsection{Participant Recruitment} \label{subsec:participants}

We recruited participants through universities and broader academic networks. Recruitment relied on snowball sampling, focusing on experienced researchers, including PhD students, postdoctoral fellows, professors, and research scientists. Participants who signed up were asked to provide information about their research interests, years of experience, scholarly background, and prior LLMs uses for research. We then applied a pre-screening procedure to ensure that participants were active researchers with sufficient expertise. Specifically, we checked their google scholar profiles for articles published in 2024-2025 and confirmed their active use of LLMs in research as marked in their pre-survey responses.

For the statistical comparison among the three control groups, a prior power calculation (based on a 3-group ANOVA test) indicated that a total 160 participants would be required to detect a medium effect size. However, recruiting this number was not practically feasible for an intensive, remote-based mixed-method study. \citet{Caine2016-pz} analyzed previous CHI user studies and reported that the median number of participants was 18. Guided by this number, we set our target sample size to $N=54$, allocating $n=18$ participants to each condition. While this number is also powered to identify large effect size (based on a post-hoc power calculation), we prioritized logistical feasibility and depth of a mixed-method approach. Quantitative results were further validated and contextualized by the rich qualitative data collected during semi-structured interviews.

Our recruited 54 researchers include 40 PhD students, 7 postdoctoral fellows, and 7 professors or research scientists. The sample consisted of 29 women and 25 men. 47 participants were based in the US, while the remaining 7 were from other regions (e.g., from East Asia and Europe). 

Our participant pool covered diverse research disciplines that roughly spanned three areas: 16 from the natural sciences and engineering (e.g., chemistry, biology, physics, materials science, medicine); 21 from the computing and information sciences (computer science, information science, and HCI); and 16 from the social sciences and humanities (political science, education, sociology, business, media and communication, art, and related fields). We report more details in Appendix \ref{appendix:demographics}.

We conducted the user study via Zoom. Participants used our system on their own computers and shared screens with us to complete the task. Surveys were administered through Qualtrics, and interviews were recorded on Zoom. Upon completion, each participant received a \$40 gift card as compensation. Recruitment and data collection took place over approximately two months from mid-June to early-August 2025. Our study was reviewed and approved by our organization's Institutional Review Board (IRB).

\subsection{Data Collection and Analysis} \label{subsec:analysis}

As a mixed-methods study, we collected both quantitative and qualitative data for analysis. Below, we outline measurements and analyses used for each.

\subsubsection{Quantitative Measurement and Analysis} \label{method:quant}
We collected participant self-reported creativity and effort scores, their interaction logs, and perceived idea ownership during the task.

\textbf{Perceived creativity support and effort.} We used the Creativity Support Index (CSI) \citep{Cherry2014-xs} and the NASA Task Load Index (NASA-TLX) \citep{Hart2006-mj, Bolton2023-cl} as dependent variables to quantify researcher perceived creativity support and effort spent. These measures were collected through a post-task survey. While prior studies employed similar instruments, they commonly used simplified versions or reported aggregated scores \citep{Liu2024-he, Liu2024-vc, Pu2025-us}. To comprehensively examine the creativity–effort trade-offs, such as how different creativity support dimensions may correlate with specific effort type, we analyzed both overall and fine-grained patterns \citep{Bolton2023-cl}.

We began by using aggregated CSI and NASA-TLX scores to assess the overall creativity–effort trade-off across control levels. The aggregated CSI scores were weighted outcomes, calculated based on participant preferences for different creativity support dimensions \citep{Cherry2014-xs}. We then analyzed individual CSI dimensions with specific effort type to assess the correlation in greater depth and whether it differed in varying control conditions. 

\deleted{For statistical testing,}  Given the differing data types and distribution characteristics, we used both parametric and non-parametric tests. We first conducted Shapiro–Wilk tests to assess normality; where normality was satisfied, we applied parametric tests, and where it was violated, we used non-parametric alternatives. We applied the Kruskal–Wallis rank sum test to assess whether reported scores varied across conditions. For correlation analysis, we applied Spearman's rank correlation to assess whether creativity and effort individually changed cross conditions. This treats control levels as ordered independent variables. We then applied Pearson correlation to analyze the relationship between creativity and effort within each condition.

\textbf{Behavioral metrics.} 
The log data includes two parts: 1) the number of standard agent-initiated actions, which were common across conditions; and 2) the number of customized functions used, which only available in certain conditions.

We used parametric statistical testing to examine whether there were significant differences in behavioral metrics across control levels. Since some behavioral metrics were intentionally disabled to differentiate the control levels (as described in Section \ref{subsubsec:UI}), we applied different statistical tests depending on whether the metrics were comparable across three conditions or only two. For metrics available under all conditions, we conducted a one-way ANOVA followed by Tukey's HSD post-hoc analysis when the ANOVA showed significance. For metrics available under only two conditions, we conducted t-tests for independent-samples. Shapiro–Wilk tests confirmed that the normality assumption was satisfied for both the three-condition and two-condition comparisons.

\textbf{Perceived Human-AI contribution and ownership of final proposals.} 
In the post-task survey, participants assigned percentages of researcher vs.\ AI contributions to the final research proposal, ensuring the total equaled 100. They then classified the proposal as `Human Work,' `AI Work,' or `Co-created Work.' Importantly, the ownership question measured participant personal beliefs about responsibility or credit, without reference to any relevant philosophical frameworks or legal statues in regard to liability, intellectual property, etc. This framing naturally can be expected to influence participant responses, and we further discuss this in Section \ref{Discuss:limitations}.

To explore potential patterns between contribution assignments and ownership classifications, we mapped the reported human–AI contributions to the corresponding ownership types. Due to the limited sample sizes across conditions and categorical work types, we did not perform significance testing on these data.

To ensure data quality, we filtered outlier responses using the {\em Isolation Forest} algorithm for anomaly detection \citep{Liu2012-tv} based on the post-task survey score distribution. Three data points were identified as extreme ratings: two unusually low and one unusually high. This resulted in 51 valid data points, with 17 for each control level. We report our quantitative results in Section \ref{subsec:quantitative_results}.

\subsubsection{Qualitative Analysis} \label{method:qual} Semi-structured interviews were conducted after participants completed the task and post-task surveys. Each interview was audio-recorded with participant consent, transcribed verbatim, and de-identified using numerical participant codes prior to analysis.

\textbf{Interview design.} Our interview focused on how participants perceived trade-offs between creativity support and effort, as well as their sense of ownership and contribution (See Appendix \ref{appendix:interview questions} for interview details). These qualitative insights extend and contextualize the quantitative analysis. We developed the interview protocol iteratively based on pilot testing feedback. 

\textbf{Thematic coding.} We conducted a thematic analysis using \added{the collaborative data analysis} to reach consensus among researchers, proposed by \citet{richards2018practical}. This involved multiple rounds of transcript review, coding, theme identification, and iterative revisions to ensure agreement. \added{Our thematic coding process includes three stages.} First, two authors independently coded subsets of the interview transcripts, with each analyzing approximately 13 transcripts. We then compared recurring patterns and shared codes, identifying initial themes related to: 1) when the system was effective vs.\ where it fell short for creativity support and 2) what effort researchers ideally wanted to contribute and felt obliged to do. By integrating themes from both parts, we consolidated trade-offs between creativity support and effort. In the second stage, each author coded the remaining transcripts from their assigned subset (approximately 13 additional interviews each). After aggregating insights from all transcripts, three overarching themes emerged: 1) effort on the creative outcome, 2) effort on the creative process, and 3) sense of ownership. In the final stage, one author applied the finalized themes and associated codes to the entire dataset. 
We present these themes in Section \ref{subsec:qualitative_results}.


\section{Findings from Quantitative Data Analysis} \label{subsec:quantitative_results}

In this section, we begin by presenting participant behavioral metrics to illustrate general use patterns across three control levels (Section \ref{quant:behavioral}). We then examine the creativity-effort trade-offs for both overall and fine-grained results (Section \ref{quant:tradeoffs}). Finally, we report participant assigned human-AI contributions and sense of ownership of the research proposals (Section \ref{quant:ownership}).

\subsection{Behavioral Patterns across Control Levels} \label{quant:behavioral}

\added{From the behavioral analysis, we found that participants generated a similar number of seed ideas and final proposals across all control levels, but their strategies for developing these seed ideas into complete drafts diverged substantially: Low-control users relied primarily on prompting full proposals, Medium-control users focused more on iterative section-level refinement, and Intensive-control users made greater use of fine-grained co-editing and critique features. We present statistical details below.}

Table \ref{tab:behavioral_metrics} shows that participants created a similar number of unique proposals and seed ideas, with no significant differences across control levels. Significant differences exist only later in the proposal development process.
For example, the metric of ``prompt\_full\_proposals'' shows statistically significance among the three control levels, $F(2,48)=6.39, p=0.004$. Based on Tukey's HSD post-hoc analysis (Appendix~\ref{appendix:post-hoc}), participants in the Low condition used this feature more often than those in the Intensive ($M_\text{diff}=6.15$, $p=0.0049$) and the Medium condition ($M_\text{diff}=5.48$, $p=0.018$). Additionally, the Medium-condition participants used ``revise\_full\_ proposals'' more often, though this difference was not significant. For ``select\_improvements'' for proposal revisions, there was a significant overall effect, $F(2,48)=4.33, p=0.019$. Post-hoc analysis revealed that participants in the Medium condition selected more improvements than those in the Low condition ($M_\text{diff}=-3.87$, $p=0.016$). Differences between Medium and Intensive conditions were not significant.

\begin{table*}[htbp]
\centering
\begin{tabular}{@{}lcccc@{}}
\hline
\textbf{Types / Metrics} & \textbf{Intensive} & \textbf{Medium} & \textbf{Low} & \textbf{Significance Testing} \\
\hline
\multicolumn{5}{l}{\textbf{\hspace{-0.5em}Metrics shared across all control levels}} \\
\# Unique\_research\_proposals & 1.58 (0.77) & 1.88 (1.20) & 2.19 (1.11) & $F(2, 48) = 1.52, p = 0.22~~$ \\
\# Generate\_seed\_ideas & 6.76 (6.47) & 5.02 (3.65) & 6.31 (4.31) & $F(2, 48) = 0.53, p = 0.59~~$ \\
\# Prompt\_full\_proposal & 5.64 (3.74) & 6.31 (4.38) & 11.79 (7.74) & \textbf{$F(2, 48) = 6.39, p = 0.004^{**}$} \\
\# Revise\_full\_proposal & 3.05 (1.73) & 4.31 (2.18) & 3.78 (2.96) & $F(2, 48) = 1.31, p = 0.28~~$ \\
\# Select\_improvements & 6.08 (4.27) & 7.34 (4.17) & 3.47 (2.60) & $F(2, 48) = 4.33, p = 0.019^{*}$ \vspace{0.5em}\\
\hline
\multicolumn{5}{l}{\textbf{\hspace{-0.5em}Specific metrics shared between intensive and medium versions}} \\
\# Query\_searches & 4.65 (4.49) & 7.36 (6.34) & -- & $t(32) = -1.48, p = 0.149$ \\
\# Select\_papers & 11.03 (18.06) & 25.01 (34.50) & -- & $t(32) = -1.54, p = 0.134$ \\
\# Prompt\_proposal\_section & 7.14 (4.08) & 11.42 (6.82) & -- & $t(32) = -2.29, p = 0.028^{*}$ \vspace{0.5em}\\
\hline
\multicolumn{5}{l}{\textbf{\hspace{-0.5em}Specific metrics only in intensive version}} \\
\# In-line\_text\_prompting & 5.72 (6.59) & -- & -- & -- \\
\# Save\_edits & 2.47 (3.83) & -- & -- & -- \\
\# Provide\_feedback & 9.57 (8.08) & -- & -- & -- \\
\# Request\_more\_critiques & 2.22 (2.83) & -- & -- & -- \\
\# Customize\_improvements & 0.59 (1.02) & -- & -- & -- \\
\hline
\end{tabular}

\vspace{0.3em}
\caption{Behavioral metrics across different control levels over agents (Low, Medium, Intensive). Values are reported as means with standard deviations based on how many functions were used per hour. Different statistical tests were used: one-way ANOVA for comparisons across all three conditions, and independent-samples t-tests for comparisons between two conditions. The significance levels are represented as p-value, 0 < ‘***’ < 0.001 < ‘**’ < 0.01 < ‘*’ < 0.05}
\label{tab:behavioral_metrics}
\end{table*}

\begin{table*}[t]
\centering
\begin{tabular}{@{}p{3.5cm}ccc@{}}
\hline
\textbf{Control levels} & \textbf{CSI\_weighted} & \textbf{CSI\_unweighted} & \textbf{Task\_load} \\
 & Median / Mean / SD & Median / Mean / SD & Median / Mean / SD \\
\hline
Low   & 59.17 / 59.65 / 18.47 & 69.17 / 66.79 / 11.90 & 54.29 / 50.42 / 13.05 \\
Medium    & 53.67 / 53.13 / 17.39 & 64.25 / 63.64 / 11.87 & 54.29 / 52.61 / 12.42 \\
Intensive & 63.87 / 63.20 / 21.77 & 72.50 / 71.51 / 15.17 & 54.29 / 51.60 / 13.57 \\
\hline
Kruskal-Wallis ($\chi^2$, df, p) & 2.08, 2, 0.353 & 3.735, 2, 0.155 & 0.251, 2, 0.882 \\
\hline
\end{tabular}
\vspace{0.3em}
\caption{Aggregated CSI and task load scores across three control levels: aggregated CSI scores exhibited a non-linear trend: highest in the Intensive condition followed by the Low and poorest in the Medium. No significant differences were found.}
\label{tab:CSI_task_overall}
\end{table*}

For metrics available only in the Intensive and Medium conditions, Medium-condition participants performed more ``query searches'' and ``selected papers'' than those in the Intensive condition, although these differences were not statistically significant. Additionally, Medium-condition participants used ``prompted proposal section'' more often to revise individual proposal sections, resulting in significant difference ($t(32)=-2.29$, $p=0.028$). For metrics unique to the Intensive, such as ``in-line text prompting'', ``save edits,'' ``provide feedback,'' ``request more critiques,'' and ``customize improvements,'' participants engaged with these features at varying levels. In particular, ``in-line text prompting'' and ``provide feedback'' were used more frequently than the other functions (both over 5 times per hour while others less than 3 in average).

Based on the behavioral metrics, we found control levels did not significantly affect the number of proposals or seed ideas participants generated. This may reflect the pre-existing ideas participants already had in mind, or possibly the specific time allowed to create ideas during our task. However, different control levels influenced how they engaged with the creative process to develop and refine proposals. We summarize below:

\begin{description}
    \item[Low]Participants prompted the system to generate full proposals more frequently as the only way to iterate drafts.
    \item[Medium]Participants iteratively selected more improvements and prompted the system more often to generate individual proposal sections, reflecting increased patterns in iterative refinement of decomposed sections.
    \item[Intensive]Participants utilized advanced, fine-grained features, especially in-line text prompting and providing feedback for agent critiques, indicating deeper user engagements over the co-editing and critiquing process.
\end{description}


\subsection{Creativity Support and Effort Changes} \label{quant:tradeoffs}

In this section, we first present aggregated CSI and NASA-TLX scores to examine how perceived creativity support and effort change separately across different conditions. We then analyze correlations between overall and individual CSI dimensions and effort types to explore fine-grained trade-offs and whether it differs across conditions. 

At an overall level, we found that creativity support followed a non-linear rating pattern across control levels (Section \ref{quant:overall changes}): highest in Intensive, intermediate in Low, and lowest in Medium, while effort scores remained similar, with no statistically significant differences in both measures. Fine-grained analysis, however, revealed notable variation across individual creativity support and effort dimensions (Section \ref{quant:fine-grained changes}): The Medium condition exhibited the strongest negative trade-offs, with high human effort associated with low creativity support. In the Low condition, trade-offs were moderate, particularly where exploration-related creativity support even increased with effort. The Intensive condition maintained a more balanced, favorable relationship between effort and perceived creativity support.

\begin{figure*}[t]
    \centering
    \includegraphics[width=0.9\textwidth]{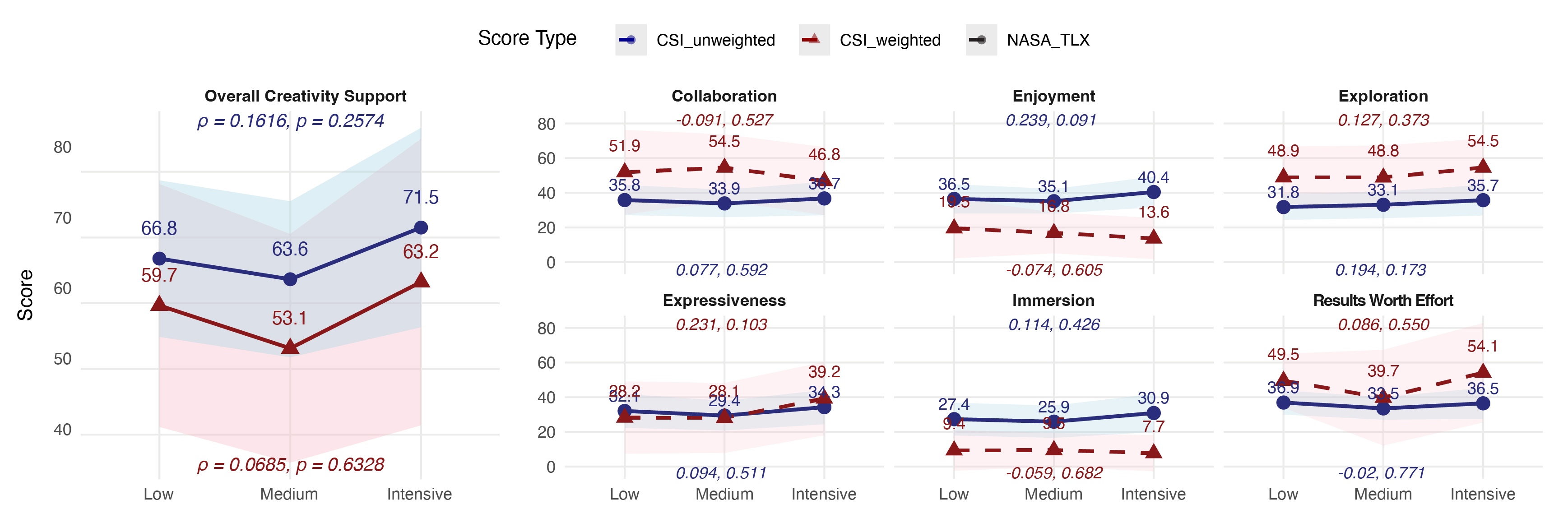}\\
    \includegraphics[width=0.9\textwidth]{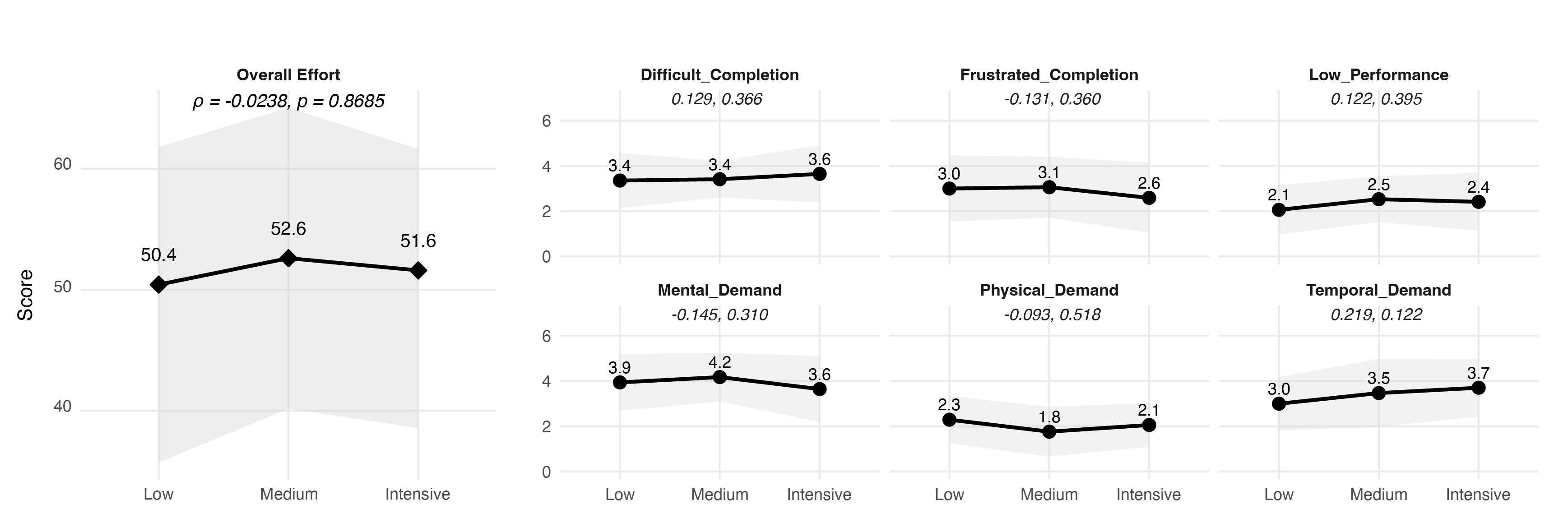}
    \caption{Changes in creativity support (weighted and unweighted scores) and effort measures across three control levels. The top panels present CSI scores. The bottom shows NASA-TLX scores. The left side of each panel displays aggregated scores, and the right break down scores by individual dimension. Spearman correlation coefficients ($\rho$) and corresponding p-values are indicated along the line.}
    \label{fig:creativity_effort_changes}
    \Description{Changes in creativity support and effort measures across the three steering levels}
\end{figure*}

\begin{figure*}[t]
    \centering
    \includegraphics[width=0.75\linewidth]{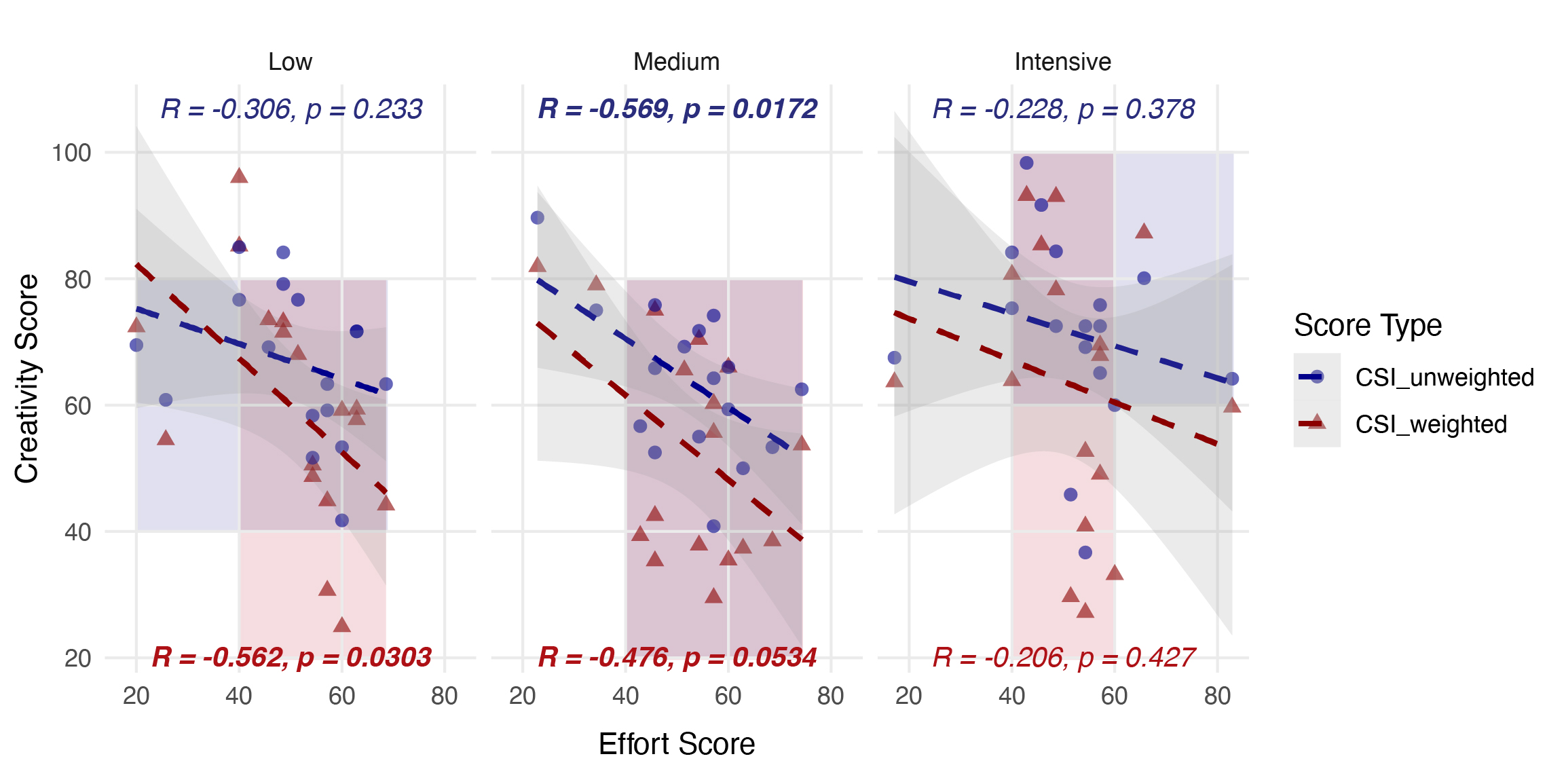}
    \caption{Correlations between overall creativity support (weighted and unweighted scores) and effort scores across three control levels. Statistically significant correlations are marked, and the highlighted rectangle indicates where nearly 80\% of data points (14 over 17 participants) are distributed. Pearson correlations were used given the parametric nature of both variables.}
    \label{fig:creativity_effort_tradeoff}
    \Description{Correlations between overall creativity support (weighted and unweighted scores) and effort scores across the three control levels.}
\end{figure*}


\subsubsection{Non-Linear Trends in Overall Creativity Support and Effort Changes.} \label{quant:overall changes}
As shown in Table \ref{tab:CSI_task_overall} and Figure \ref{fig:creativity_effort_changes}, aggregated CSI scores exhibited a non-linear trend: 
CSI scores were highest in the Intensive condition 
($M_\text{weighted} = 63.2$, $M_\text{unweighted} = 71.5$), 
    followed by the Low 
($M_\text{weighted} = 59.7$, $M_\text{unweighted} = 66.8$), 
and poorest in the Medium 
($M_\text{weighted} = 53.1$, $M_\text{unweighted} = 63.6$). 
These differences were not statistically significant
($\chi^2(2) = 2.08$, $p = 0.353$ for weighted; 
$\chi^2(2) = 3.735$, $p = 0.155$ for unweighted). The correlation between CSI scores and control levels were also weak and nonsignificant 
($r = 0.0685$, $p = 0.6328$ for weighted; 
$r = 0.1616$, $p = 0.2574$ for unweighted).

When comparing the weighted and unweighted scores by creativity support dimensions, weighted scores were higher in \textit{collaboration}, \textit{exploration}, and \textit{value} on average, indicating these dimensions were considered more important by participants when using our system to develop research proposal. In contrast, weighted scores were lower than unweighted scores in \textit{enjoyment} and \textit{immersion} on average, suggesting these aspects were viewed as less critical.

Effort scores showed minimal variation across control levels.
The Medium condition recorded the highest effort score
($M = 52.6$), 
followed by the Intensive 
($M = 51.6$), 
and the Low condition 
($M = 50.4$). 
These differences were not significant
($\chi^2(2) = 0.251$, $p = 0.882$), 
and correlations were negligible 
($r = -0.0238$, $p = 0.8685$). 


\subsubsection{Fine-grained Trade-Offs between Different Creativity Support and Effort Dimensions.} \label{quant:fine-grained changes}
Figure~\ref{fig:creativity_effort_tradeoff} shows that perceived creativity support decreased generally as effort rises across all control levels. However, the strength of this relationship varied. 

In the Low condition, the negative correlation was significant for the weighted score ($R = -0.562$, $p = 0.0303$) but not for the unweighted score ($R = -0.306$, $p = 0.233$). In the Medium condition, both weighted and unweighted scores showed strong negative correlations, with the weighted score reaching significance ($R = -0.569$, $p = 0.0172$) and the unweighted score approaching significance ($R = -0.476$, $p = 0.0534$). However, the Intensive condition showed a more favorable balance between creativity support and effort, where the negative trend was weaker and non-significant for both weighted ($R = -0.206$, $p = 0.427$) and unweighted ($R = -0.228$, $p = 0.378$). Compared to the dispersed patterns in the Low and Medium conditions, the Intensive condition showed a more centralized distribution (highlighted rectangle area in Figure \ref{fig:creativity_effort_tradeoff}), with unweighted scores clustering in areas of high effort and high creativity support.

Post-hoc analysis on each specific creativity and effort dimension revealed more nuanced patterns. The perceived creativity support relative to effort differed substantially across individual dimensions, with some even showing positive trends where creativity support increased alongside certain types of effort. More detailed descriptions are presented in Appendix \ref{appendix:fine-grained_tradeoffs}. We provide a brief summary:
\begin{description}
    \item[Intensive]This condition showed increased scores on \textit{collaboration} and \textit{expressiveness} for creativity support, especially when participants perceived \textit{tasks as challenging}.
    \item[Medium]This condition showed the lowest creativity support, with the strongest negative trade-offs across all effort type.
    \item[Low]This condition showed increased scores on \textit{exploration} for creativity support, particularly when participants experienced higher \textit{mental}, \textit{physical}, and \textit{temporal} demand.
\end{description}


\subsection{Human-AI Contributions and Ownership} \label{quant:ownership}
\begin{figure*}[t]
    \centering
    \includegraphics[width=0.7\linewidth]{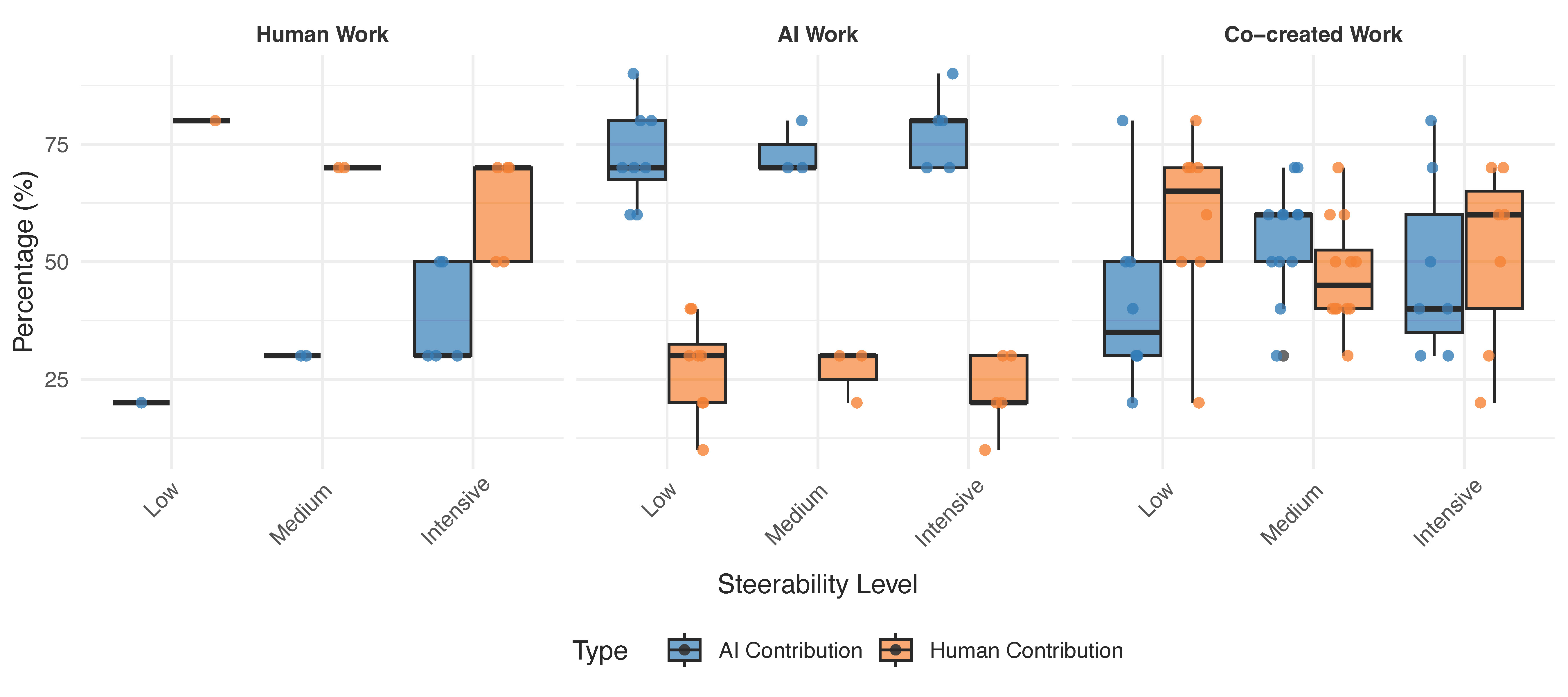}
    \caption{Distribution of human and AI contributions and perceived idea ownership across three control levels. Participants estimated the percentage of each contributor input and categorized final research proposals as human-generated, AI-generated, or co-created.}
    \label{fig:idea_contribution}
    \Description{Distribution of human and AI contributions and perceived idea ownership across three control levels.}
\end{figure*}

By mapping contribution percentages onto ownership types (Figure~\ref{fig:idea_contribution}), we found that \deleted{distinct patterns in how participants perceived idea ownership relative to the level of control} \added{participant perceptions of idea contribution and ownership varied by type: ``Human Work'' and ``AI Work'' were consistently attributed to one party regardless of control level, while ``Co-Created Work'' reflected a more balanced negotiation of contributions between humans and AI.}

Only 15.7\% (\textit{N=8}) of participants classified ideas as `Human Work.' They consistently attributed higher contributions to themselves across all control levels, with just one participant assigning 50\% human contribution in the Intensive condition. This suggests they viewed their original input as the primary source of the idea.
31.4\% (\textit{N=16}) of participants classified ideas as `AI Work,' consistently assigning higher contribution percentages to the AI regardless of control level. This indicates that even with more control, the AI is still seen as the dominant contributor. 

Only 52.9\% (\textit{N=27}) of participants classified ideas as `Co-Created Work.' Attribution varied across conditions but were more balanced than in other ownership types. In the Low condition, four participants assigned higher contributions to themselves, one to the AI, and two equally. In the Medium, five attributed more to the AI, three to themselves, and three equally. In the Intensive, four assigned higher contributions to themselves, two to the AI, and one equally.



We found that when participants classified ideas as Co-Created Work, they perceived a more balanced distribution of contributions between humans and AI. Unlike Human Work or AI Work, where attribution was consistently skewed toward one side, co-created ownership reflected an active negotiation between human and AI. In contrast, judgments of Human Work and AI Work appeared to be driven by deeper factors, leading participants to consistently assign higher contributions to one party across control levels.

Our qualitative findings suggest that this pattern stems from how participants prioritize idea originality and execution effort, and their perceptions on who exert main controls over the final outcomes. We report these details in Section~\ref{qual:sense of ownership}.

\section{Findings from Qualitative Data Analysis} \label{subsec:qualitative_results}

Based on our thematic analysis of the interview data (Section \ref{method:qual}), we identified two trade-off patterns involving creativity support and effort: the system influence on researcher effort spent on creative outcomes (Section \ref{qual:creative outcomes and effort tradeoffs}) and the creative process (Section \ref{qual:creative process and effort tradeoffs}). We identified three key factors that shaped how participants attributed contributions and ownership: the originality of ideas, the effort required for execution, and the recognition of human–AI work divisions (Section \ref{qual:sense of ownership}).


\subsection{Effort Spent on Creative Outcomes} \label{qual:creative outcomes and effort tradeoffs}

\subsubsection{Low Effort in Idea Generation but with Mixed Quality} \label{qual:low-effort idea generation}
Our research ideation system allowed participants to put in relatively less effort than writing the proposal on their own, but the resulting outcomes showed inconsistency in idea quality. 

On the positive side, the low-effort enabled rapid exploration of many ideas. Low-condition participants particularly acknowledged this value. 
For example, P50 (Low, Research scientist) noted: ``It was rewarding to have a nicely generated proposal idea quickly, productivity-wise...enabled me to express ideas that I hadn't really had time to flesh out.'' P39 (Low, PhD-Student) appreciated the scaffolding feature, enabling researchers to move from scattered ideas to concrete starting points: ``It gives you a structure even if it's not perfect, a scaffolding strategy where to start and move forward to the next stage.'' 

However, the ease of idea generation leads to concerns around inflated expectations of research capacity. P50 who appreciated the productivity benefits pointed out that: ``This tool made it too easy to come up with research ideas...I would end up using it for more and more ideas...but it's not realistic that I could actually perform that amount of research.''

Additionally, participants provided mixed review of the idea quality. For example, some initially found the drafts appealing, but later questioned whether they were feasible. This concern was consistent across all conditions. 
For example, P02 (Intensive, PhD-Student) questioned that: ``They show some interesting things, looks very fancy, but is not actually what we can do.'' Similarly, P42 (Medium, Postdoc) noted: ``I was looking for inspiration, to see what's possible. 
In that sense, the system did a good job
but I don't think they're now workable yet.'' P41 (Low, PhD-Student) shared similar feelings, saying: 
``Some ideas seem promising but it can be a completely different story once you start working in the lab.''

Furthermore, even when ideas presented a workable path, some participants viewed them as a compromise on novelty. As P03 (Medium, PhD-Student) noted: ``I was surprised that the tool really did know which field it should be referring to and citing.'' This alignment with disciplinary norms often reinforced existing ideas rather than challenged them. P10 (Intensive, PhD-Student) described as ``repetitive and not really novel,'' and P15 (Low, PhD-Student) similarly reflected, ``When it comes to creativity in terms of research methods, framework of the question and design, I think it is largely similar to prior research...I wonder how I can be more creative?''


\subsubsection{High Effort Directed toward Ideas Verification} \label{qual:high-effort idea verification} 

While the system reduced the effort required for initial ideation --- by selecting a seed idea from the \textit{Ideator} and, with support from the \textit{Writer} and \textit{Evaluator}, developing it into a full proposal --- it increased downstream human work, including verifying citations, evaluating methodologies, and assessing implementation feasibility. This shift in effort introduced cognitive overhead, as participants were responsible for validating content in drafts without having been involved in the underlying reasoning.

For example, participants wanted to understand the logic behind AI citations. P32 (Intensive, PhD-Student) explained: ``It would be more helpful if the system not only gives the references but also provides explanations for why those references were included.'' Beyond citations, participants sought deeper methodological elaboration. P33 (Intensive, Professor) emphasized the importance of expanding methodological details to contextualize why an AI-suggested method could be considered novel: ``If I were to move forward with one of these studies and call it creative and novel, I'd want [the system] to expand more on how it works and why it is novel.'' 

Idea verification was especially challenging when participants lacked domain expertise in suggested methodologies. For example, P45 (Intensive, PhD-Student) described: ``It asked me to use social network analysis, and I wasn't able to evaluate whether this method is actually viable and feasible for the research question.'' As a result, participants naturally adopt, as P39 (Low, PhD-Student) described, ``a reviewer mindset'' to passively take responsibility: ``You have to double check because the things that you're going to submit to the journal or the conferences, this becomes your responsibility.''

When participants had sufficient domain knowledge, their verification efforts extended to prompt writing, incorporating more literature-based details to guide feasible implementations. However, sentence-level fine-tuning did not always yield precise revision, heavily depending on the retrieval context. For example, P51 (Intensive, PhD-Student) successfully prompted the inclusion of technical specifications: ``I added `could you please include the length [of a bio-material]?'... It included five to ten kilobase pairs, within the range my PI suggested.'' In contrast, other attempts produced unintended rewrites. P49 (Intensive, PhD-Student) said, ``I highlight a sentence and ask the AI to give me more details but then it replaces the whole paragraph...lose crucial information that was there before.'' This varying idea quality was particularly evident in the Intensive condition. Participants noted that, while the interface offered finer-grained controls, achieving the desired results required substantial effort to monitor and correct unintended changes.


\subsection{Effort Spent on the Creative Process} \label{qual:creative process and effort tradeoffs}

\subsubsection{Low Effort in Writing to Converge Thoughts, but Loss of Cognitive Engagement} \label{qual:low-effort converging ideas}

Participants frequently highlighted that the system reduced the effort required for writing. Research ideation involves integrating concepts, identifying connections, and uncovering gaps, and AI helps by directly generating coherent drafts that assemble these elements. However, this comes at a cost: cognitive erosion and the loss of emotional fulfillment that researchers experience when tackling these tasks themselves.



P43 (Intensive, PhD-Student) highlighted the trade-off between reduced writing effort and diminished cognitive engagement: ``The tool is good at finding literature and outlining big-picture ideas...But actually writing things out by hand helps me think about them more deeply.'' This could lead to passively accepting AI-generated outputs, as they further explained: ``It made it easier to just accept what it generated as truth. It cites papers, sounds reasonable. But I hadn't worked through the reasoning myself, I lost some agency in the research process.''
P12 (Low, Research Scientist) directly characterized this as a risk of ``cognitive erosion --- skills without the learners or the experts being aware that their skills are eroding.''  

The effortful process of wrestling with ideas through writing, specifically what our systems were designed to minimize, proved to be intellectually and emotionally rewarding for human researchers. For example, P42 (Medium, Postdoc) described: ``A lot of the time, writing the literature review or the introduction can feel tedious. But you end up discovering a lot, not just about the topic, but also about yourself.'' Similarly, P37 (Medium, PhD-Student) considered writing research proposals is a path for intellectual satisfaction by ``coming up with your own research questions and your own approach to problem-solving.'' P47 (Medium, Postdoc) also echoed ``these are what something I want to lead, not the AI.''
P12 (Low, Research scientist) drew an analogy to aviation: ``Airplane pilots also use autopilots but they must be prepared to step in by going through rigorous training and retraining to hone their cognitive abilities.'' In the same spirit, P12 cautioned that junior researchers should use the system carefully, emphasizing that without sufficient preparation, their research directions might be constrained by AI outputs rather than allowing them to think creatively.




\subsubsection{High Effort Directed toward Understanding the System} \label{qual:high-effort understanding system}

Although our system was designed to streamline research ideation, participants found that using the system required increased attention to navigate the system, which disrupted immersion in the ideation process. This occurred across control levels. 

Rather than focusing on creative thinking, their effort was redirected toward monitoring the interface, managing structural organization, and ensuring the presentation of outputs. For example, P50 (Low, Research scientist) appreciated having all functions in one place, but also acknowledged additional mental effort: ``I enjoyed having a little bit less control...This AI solves the problem of going to many different places to do tasks by combining them in one window. However, you still need to manage and coordinate everything mentally.'' Similarly, P42 (Medium, Postdoc) highlighted the challenge of tracking revisions based on criteria and incorporating suggested improvements: ``I kept clicking...and removed the novelty and feasibility...so I only included those criteria and removed everything but then I had to look back on the different versions to really understand how that request changed the proposal.'' P16 echoed similar feelings, arguing that "I gave the AI keywords, and it gave me a whole page to read."

\subsection{Sense of Ownership} \label{qual:sense of ownership}
Across different control levels, we identified two recurring factors---idea originality (Section \ref{qual:originality}) and writing execution effort (Section \ref{qual:execution})---that influenced how participants assigned contributions and ownership. The key determinant of ownership was their perception of who held control. When participants viewed either the human or the AI as exerting greater control, they attributed ownership accordingly; when they recognized complementary roles between the two, this led to shared ownership (Section \ref{qual:work division}).  

\subsubsection{Idea Originality} \label{qual:originality}
When participants felt they came up with the original ideas, they claimed ownership, even if AI produced most of the draft. 
For example, P07 (Intensive, PhD-Student)
explained: ``The most important thing for the proposal is that the idea comes from myself. AI is just polishing it...
so 70\% to me, 30\% to AI''. P23 (Medium, Research scientist) also noted, ``Some specific details were totally by AI, but broad strokes, top-level organization, and the original idea comes from me.'' 
Even for some participants who used the Low condition also claimed ownership, noting that ``Idea-wise, 90\% is from me'' (P39, Low, PhD-Student). 

The originality also comes from changing critical elements of generation, such as making the proposal more feasible. P34 (Low, PhD-Student) explained: ``By the time I fix the impact that I want, the method I want to use... those things are very important in research... you see a sense of direction is already there, that's why I give myself a high percentage.''

In contrast, when participants contributed idea originality little beyond keywords, ownership shifted more toward the AI, though this was relatively rare among participants. 
P06 (Low, PhD-Student) admitted: ``I gave very few keywords... didn't even write a sentence...the seed idea generated by AI was very good.'' Yet in other cases, when participants acknowledged AI also contributing to the original idea, they considered the outcome as co-created work (we elaborate this specific co-created case in Section \ref{qual:work division}). 

In conclusion, these participants grounded their sense of ownership in who originally had the idea. Even when AI contributed significantly by refining or rephrasing those ideas, participants maintained intellectual agency, viewing themselves as the primary source of research ideation. 

\subsubsection{Execution Effort} \label{qual:execution} Another factor is execution, which refers to who ultimately transforms ideas into complete proposals. Many participants viewed AI as the primary author because it performed most of the writing. While they acknowledged their own inputs and revisions, participants tended to assign ownership based on the execution effort, especially when AI outputs were of high quality.

Intensive-condition users, who had greater control over the process, emphasized the role of execution. P22 (Intensive, PhD-Student) explained, ``I think the high-level idea was mine, but all the technical details came from AI, that's the most important on how you would actually execute that. So I chose 80\% for AI.'' P51 (Intensive, PhD-Student) similarly explained that they contributed only seed keywords, while the AI carried out the substantive work in a ``well-structured content'', deserving most ownership.
P25 (Intensive, PhD-Student) even reported that nearly all effort came from the model, yet hesitated to cede full ownership to AI. 
``Aside from the initial seed idea, the text was almost entirely generated by AI about 90\%. Still, I hesitate to say the work is fully AI-owned because I provided the prompts.''

The execution effort involves not only drafting the initial proposal but also iteratively revising it based on evaluation. 
Some participants viewed this stage as an opportunity to regain a sense of contribution and ownership.
For example,
P11 (Medium, Professor) described, ``It can write almost everything, but without human involvement in where the parts were missing, the quality of the writing would remain at a very low level. I think our role is translating now to more of reviewing, editing, and giving comments to AI.'' P10 (Intensive, PhD-Student) also emphasized, ``AI cannot judge scientific soundness...so definitely reviewing should not be done by AI.'' However, occasionally, some participants also defer this contribution to AI 
as P10 (Intensive, PhD-Student) noted,
``I didn't do much, just gave a keyword that generated potential research ideas. The AI handled most of the literature review... Even for the evaluation, like feasibility and novelty, was done by the AI.'' 

These execution focus helps explain why the Intensive-control participants, despite having the greatest degree of control, still attributed primary ownership to the AI. 
Although the Intensive condition provides granular co-editing and evaluation features that make AI automated execution more visible, this transparency does not diminish its perceived ownership. These participants perceived that AI carried out most of the writing work. 


\subsubsection{Acknowledging Clear Work Division Contributes to Shared Ownership.} \label{qual:work division}
When participants recognized complementary roles with the AI in proposal development, they regarded the draft as \textit{Co-Created}. Humans contributed original ideas, defined the scope, and provided feedback, while the AI handled rapid drafting, synthesis, and sentence-level polishing. This division of labor emphasized that humans and AI played distinct but equally important roles in shaping both the process and the outcome.

Some participants articulated a sophisticated understanding of human–AI collaboration, highlighting the distinct roles each played rather than the sheer amount of content generated. For example, P11 (Medium, Professor) mentioned: "Although sentences were mostly generated by AI, the idea and direction were made by humans...It's not the quantity that AI did more, but it's about what roles humans can actually play, and how AI can support that role.''

Other participants described a more iterative and collaborative pattern in which humans and AI play distinct roles rather than comparing contribution amounts. P42 (Medium, Postdoc) explained: ``I provided AI with the essential keywords...AI elaborated on it and then I revised that elaboration...The system also generated a lot of synthesized literature, which helped me complete the overall narrative...But I was still actively shaping the idea, trimming parts that didn't fit, asking for specific framings, and guiding revisions...so 50/50 between me and the system.'' This illustrates how the iterative refinement between humans and AI, particularly through feedback features, shaped perception of co-creation. 
Even within the Low condition, participants who possessed sufficient topic knowledge and provided details in the prompt perceived the work as co-created. P41 (Low, PhD-Student) said 
``My ideas were already in progress. I had narrowed down the target proteins and molecules I wanted to focus on for developing a combination therapy.''


\section{Discussion} \label{sec:discussion}

\subsection{Integrated Insights from Quantitative and Qualitative Findings}

Our quantitative results revealed nuanced patterns in creativity support and effort. The results from qualitative data analysis help contextualize these patterns by illustrating how participants experienced effort and interaction with the system. In this section, we integrate both perspectives to address two RQs regarding how different control levels over LLM agents affect the creativity-effort trade-offs for research ideation and how they influence the division of contributions between humans and AI, and researchers' perceived ownership of final outcomes.

\subsubsection{Trade-offs between Creativity Support and Effort Spent for LLM-Supported Research Ideations}

By combining overall and fine-grained quantitative results (Sections \ref{quant:overall changes} and \ref{quant:fine-grained changes}), we found that: 1) the trade-offs between creativity support and required effort are non-linear, and 2) different levels of control reveal varying relationships between specific dimensions of creativity support and types of effort.

First, increased control did not necessarily enhance creativity support in generating ideas and proposals, and effort scores remained consistent. Qualitative findings suggest this is because the AI still took substantial effort for generating ideas (Section \ref{qual:low-effort converging ideas}) and writing proposals (Section \ref{qual:low-effort idea generation}). Participant effort was instead redirected toward verifying ideas (Section \ref{qual:high-effort idea verification}) and understanding the system (Section \ref{qual:high-effort understanding system}) regardless of control levels.


Second, although lower creativity support was associated with increased effort, both the Low and Intensive conditions still showed positive trends on certain creativity dimensions, such as \textit{exploration} in Low condition, and \textit{collaboration} and \textit{expressiveness} in Intensive condition. This is because both levels enabled participants to manage trade-offs: either by maximizing AI benefits or minimizing its drawbacks. For example, in the Low condition, participants maximized AI benefits: minimal control allowed them to quickly generate a large volume of ideas, increasing exploration despite higher mental, physical, and temporal demands (Appendix \ref{appendix:fine-grained_tradeoffs}). This improved system efficiency but came at the expense of deeper cognitive engagement (Section \ref{qual:low-effort idea generation}). In the Intensive, participants minimized AI drawbacks: the higher degree of control provided additional features to refine outputs. Collaboration and expressiveness increased despite higher perceived task difficulty (Appendix \ref{appendix:fine-grained_tradeoffs}). This partially mitigated the loss of cognitive engagement, particularly the effort involved in verifying ideas (Section \ref{qual:high-effort idea verification}), though it did not fully resolve it.
 

\added{These nuanced trade-off effects of controls highlight the need for design strategies that tailor LLM-supported research ideation to different user objectives. We discuss this further in Section \ref{Discuss:multi-objective design}.}

\subsubsection{Human-AI Contributions and Perceived Proposal Ownership}

We identified distinct quantitative patterns in how participants attributed contributions between humans and AI across control levels. Those who classified the proposal as either 'Human work' or 'AI work' consistently viewed one side as the primary contributor, regardless of control levels (Section \ref{quant:ownership}). In contrast, 'Co-Created Work' showed more varied patterns across control levels. Qualitative findings suggest that ownership judgments are driven primarily by how participants weigh idea originality (Section \ref{qual:originality}) vs.\  execution effort (Section \ref{qual:execution}). Depending on which factor they prioritize and who exerts control over it, participants consistently assign ownership in accordance with that emphasis. Shared ownership emerged when they recognized and acknowledged a clear labor division of human-AI collaboration (Section \ref{qual:work division}).

Based on the aforementioned results, we found the degree of control did not strongly influence participants attributed contributions and ownership. While some participants reported a decreased sense of ownership when using the system (similar to prior work \citep{Draxler2024-ii, Kreminski2024-pf}), others consistently expressed a strong sense of ownership. A key factor appears to be whether participants can utilize their knowledge, expertise, and, more importantly, act in accordance with the ethical sense of responsibility to use the system effectively. Reflecting on prior work \citep{Joshi2025-hp}, participants with high expertise tended to write long, detailed prompts to guide AI generation, thereby increasing their sense of ownership.

Two additional themes merit further discussion. The first underscores human creative adaptability within system constraints (see Section \ref{Discuss:adaptability in creativity}). The second concerns how we define the boundaries of shared ownership between humans and AI (see Section \ref{Discuss:ownership}).

\subsection{Tailoring Agent Controls to Diverse Researcher Values for Research Ideation} \label{Discuss:multi-objective design}

While a mixed-initiative approach is widely recognized by HCI scholars in designing LLM-supported research ideation tools (Section \ref{subsec:design_spectrum}), its underlying design principle for most recent work is often grounded in the concept of complementary performance \citep{Cimolino2022-rc}. This principle emphasizes leveraging the respective strengths of humans and AI to enhance productivity goals, such as work efficiency or efficacy, across different human–AI teaming tasks, such as decision-making \citep{Chiang2023-th}, fact-checking \citep{Nguyen2018-qn, Das2023-fg}, data analysis \citep{Chen2025-ep}, as well as scientific discovery \citep{Chan2018-of, Kang2023-yh}. However, our qualitative findings suggest that productivity is not always the primary value in creativity-oriented domains such as research ideation.

As described in Section \ref{qual:creative process and effort tradeoffs}, participants often prioritized different goals such as intellectual growth, professional development, and meaningful engagement with their ideas, even when this required cognitively demanding activities including reading, writing, and organizing notes to support deep, reflective thinking. For many scientists, these effortful processes were enjoyable. \added{Therefore,} we argue that for highly intellectual tasks like research ideation, \added{designing }mixed-initiative work decomposition should prioritize a broader set of human-centered goals, such as emotional fulfillment and personal meaning-making, rather than solely reducing cognitive load or maximizing AI automation. 

\added{By broadening the values and goals, the design space for research ideation can become more diverse and flourishing, encompassing a wider range of control features or strategies.} \deleted{These goals could balance short- and long-term objectives, guiding when to prioritize efficiency to meet short-term productivity objectives vs.\ allocating reflective effort to sustain long-term intellectual growth. Designers should emphasize (when possible) human engagement in tasks that researchers value, enabling a balance between efficiency, ownership, creativity, and learning. Given dynamic shifts between short- vs.\ long-term mindsets, tools should offer flexible solutions for these varying use-cases.} \added{For example, interest-oriented search \citep{Mei2025-wm} and facet-based filtering \citep{Radensky2024-gq} could be incorporated into large-scale exploration of AI-generated ideas. These design strategies can support short-term productivity by enabling human researchers to scan and evaluate AI-generated ideas based on their domain expertise. For long-term intellectual growth, features such as meta-cognitive prompts \citep{Singh2025-jj, Tankelevitch2024-at} or visualizations of AI reasoning landmarks \citep{Bogdan2025-yr} can be integrated into a coherent, end-to-end human–AI ideation system, such as the one we developed. These not only enable researchers to examine how ideas evolve, but also foster deeper understanding of AI knowledge, promote self-learning, and support the development of researcher own creative and analytical capacities over time.}



\subsection{Human Adaptive Creativity within Constraints} \label{Discuss:adaptability in creativity}

The non-linear patterns of creativity support and nuanced difference across control levels illustrate how system constraints shape user expectations, guide behavior, and elicit adaptation driven by human intrinsic motivation. In the Low condition, participants adapted by providing detailed prompts and iteratively refining outputs. In the Intensive condition, they directly shape outputs and refine ideas via granular controls. In contrast, the Medium condition created an ambiguous middle ground: it introduced enough controls to impose cognitive overhead but not enough to feel empowering, leaving participants uncertain about how to interact effectively with the system.

These observations suggest that humans often adapt creatively to clear constraints, finding ways to work within or around them even under strict limitations. In psychology and management research on creativity, such adaptability is recognized as a hallmark of human creative behavior, enabling individuals to navigate and innovate within defined boundaries \citep{Acar2019-lm, Tromp2023-xs}. This implies that future HCI research could move beyond framing human-AI interactions merely as a trade-off between control and freedom, as in our study, and instead consider it as a strategic system for imposing constraints that foster creativity and guide users toward innovative solutions.

\subsection{Defining Credit and Responsibility for Shared Idea Ownership between Humans and AI} \label{Discuss:ownership}

Our study measured participant perceptions of ownership, which were surveyed by their perceived contributions to the human-AI collaborative process in research ideation and proposal development. The results indicate a tension in how credit is assigned: some researchers valued originality of ideas over execution effort, while others held the reverse view or considered both equally important. These subjective perceptions of ownership might not necessarily align with legal or formal frameworks in the science community \citep{Editorials2023-hd, Bozkurt2024-ds}. Additionally, an individual feeling of ownership over an AI-assisted idea
may directly contradict established intellectual property law or liability precedent.

Our findings suggest several directions for defining ownership of human–AI co-created ideas. First, it is important to establish clear attribution guidelines in advance, as participants reported mixed credit between idea originality and execution effort (Section \ref{qual:work division}), and to ensure transparent reporting of contributions. \added{To address these tensions, incorporating design reflexivity into human–AI ideation systems may help researchers explicitly examine and negotiate contribution and authorship \citep{Cambo2022-oi, Pihkala2016-am}. For example, a public hub of AI-generated research ideas, such as recent initiative Hypogenic AI}\footnote{Hypogenic AI: \url{https://hypogenic.ai/ideahub}}, \added{can provide a venue for scientists to collectively evaluate and discuss AI-generated research. A reflexive design could further disclose how each idea was produced, such as initial human prompts, model parameters, intermediate generations, or subsequent human refinements. By making these processes visible, such reflexive scaffolds provide a more transparent and accountable basis for discussing ownership.}

\added{Additionally,} future studies should also examine how personal beliefs about ownership align with broader societal or legal norms. Prior work on patent ownership in crowdsourcing may offer useful examples \citep{Wolfson2011-to}, exploring how credit and liability are assigned under different legal or ethical contexts, such as patent inventorship, data security, or copyright. Defining these boundaries is essential for identifying potential risks when adopting such systems and for guiding designers to create collaborative human–AI systems that support fair, transparent, and legally informed attribution. Researchers and their organizations would benefit from understanding where personal beliefs about ownership align---or diverge---from societal and legal standards, which may evolve alongside new AI capabilities.



\subsection{Study Limitations} \label{Discuss:limitations}

We acknowledge several limitations, particularly regarding task context and disciplinary coverage. First, our study focused on research ideation during proposal development and was conducted over a short period, with participants spending approximately 35–40 minutes, compared to real-world proposal writing. In practice, research ideation is iterative and unfolds over longer timeframes, during which participants’ sense of ownership, agency, and perceived control may evolve. Several participants noted that more time would allow them to refine ideas, adjust prompts, or experiment further with the AI. Thus, our findings primarily reflect early-stage interactions rather than longer-term dynamics of using the system.

Second, although participants came from diverse research disciplines to enhance generalizability, this breadth may have limited depth within specific fields. Different disciplines have unique norms, methodologies, and epistemic practices that shape AI interaction. Consequently, some nuanced, field-specific insights may not have been fully captured. Future work could investigate domain-specific variations to better understand these differences.

\section{Conclusion} \label{sec:conclusion}


Our study examines how varying control levels over LLM agents influence the trade-offs between creativity support and effort in research ideation, the division of contributions between humans and AI, and researcher perceived ownership of outcomes. We developed an agentic ideation system integrating three roles---Ideator, Writer, and Evaluators---across three control levels---Low, Medium, and Intensive. By collecting use data from researchers from diverse disciplines, we found complex trade-offs in creative labor division: human effort shifts from producing original ideas to verifying ideas, final ownership emerges as a negotiated outcome between human and AI, and the degree of control over agents only partially shapes the creative process. We suggest LLM agents should empower researchers, fostering a sense of ownership over good ideas rather than reducing them to operators of an AI-driven process, while also recognizing dynamically varying researcher goals over different time horizons (e.g., meeting short-term productivity objectives vs.\ long-term goals for intellectual growth and development) may motivate flexible support with varying affordances. 

\section{Acknowledgments}

We extend our sincere gratitude to the participants from diverse academic communities worldwide, whose contributions made this research possible. This work was partially funded by CosmicAI through support from the NSF (Cooperative Agreement 2421782) and the Simons Foundation (grant MPS-AI-00010515), as well as by Good Systems (a UT Austin Grand Challenge dedicated to developing responsible AI technologies), the School of Information, and a Research Fellowship from UT Austin\footnote{The NSF-Simons AI Institute for Cosmic Origins (CosmicAI): \url{https://www.cosmicai.org/}; UT Good Systems: \url{https://goodsystems.utexas.edu/}}. Additional computational resources were provided in part by Amazon, Google, and the Texas Advanced Computing Center (TACC). The statements made herein are solely those of the authors alone and do not necessarily reflect the positions of the supporting institutions.

\bibliographystyle{ACM-Reference-Format}
\bibliography{reference}

\appendix
\section*{APPENDIX} 

\section{RAG Integration and Prompt Details for the Agentic Writing Workflow} \label{appendix:prompt details}

As described in Section \ref{subsubsec:implementation}, we adapt the enhanced RAG architecture (Figure \ref{fig:RAG_architecture}) to each agentic role, accounting for differences in complexity and tool requirements. Below, we summarize these differences:
\begin{description}  
  \item[Ideator] This role does not require \textit{zero-dense retrieval} as no specific seed idea is available at the initial stage. Thus, we use standard query rewriting to retrieve relevant academic articles as context and prompt the LLM to generate multiple seed ideas with varied research directions. This approach is inspired by \citet{Liu2024-he} breadth-first prompting strategy for divergent thinking.
  
  \item[Writer] This role requires a more flexible RAG configuration to incorporate references, including inline citations and user-specified key references, as well as generating different research proposal sections. We integrate direct retrieval of user-selected references from the search sidebar, combine them with default references, and rerank by relevance to the query. To enable inline citations, each reference is assigned with a unique ID, which the LLM appends when using the reference. An additional fact-checking LLM call is used to ensure content accuracy and help correct missing or misused references.
  
  The proposal has three subsections, literature synthesis, research goals, and study plan, so we use \textit{chain-of-thought prompting} \citep{Wei2022-pi} to guide the LLM sequentially: synthesize literature, generate research goals, then create a study plan. This approach allows users to intervene at each step without regenerating the entire proposal. Retrieved references are stored in short-term memory for reuse when prompting individual sections or using highlighting texts for LLM revisions.

  \item[Evaluator] This role employs a standard RAG structure with a modified generation step that also uses chain-of-thought prompting to first generate critiques, followed by proposed improvements. To enable automatic proposal iteration based on the \textit{Evaluator} feedback, both the \textit{Ideator} and the \textit{Writer} incorporate the \textit{Evaluator} improvements as additional prompt context to refine the research idea and revise the existing proposal.
\end{description}

We organize our prompt details based on the three agentic roles: \textit{Ideator}, \textit{Writer}, and \textit{Evaluator}.

\subsection{Prompt Details for Ideator} \label{appendix:ideator_prompt}
The \textit{Ideator} agent generates seed research ideas from given topics using RAG. It also includes idea refinement based on improvements suggested from the \textit{Evaluator}. We present these prompts below.

\begin{lstlisting}
IDEATION:
    Brainstorm {num_ideas} novel, feasible, and impactful research directions based on user provided {topic}.
    
    Brainstorming criteria:
    Novel: The research idea presents a new and original contribution to the field.
    Feasible: The research can be completed with available resources and within a reasonable timeframe.
    Impactful: The research has potential to make a significant contribution to theory or practice.
    Here is relevant academic context to inspire your ideas: {context}
    Web Results (Optional): {web_results}
    
    INSTRUCTIONS: 
    1. Each idea should point to a different research direction.
    2. Be creative and think about unexplored angles where these fields intersect.
    3. Ground your ideas in the academic literature and criteria.
    4. Each idea should be a single sentence without any punctuation, starting with a capital letter.
    5. Do not include citations in your ideas. 
    
\end{lstlisting}

\begin{lstlisting}
IDEA REVISION:
    Your task is to revise the following research idea.
    
    Original research idea: {query}
    Current research idea: {current_research_idea}
    Evaluation feedback: {criteria_text}
    
    Based on this feedback, please generate a new and simple one-line research idea. 
    IMPORTANT: 
    1. The idea should be like single research title without any punctuation.
    2. Do not include any citations or references.
    
    New research idea:
\end{lstlisting}

\subsection{Prompt Details for Writer} \label{appendix:writer_prompt}

The \textit{Writer} agent develops seed ideas into full research proposals, comprising three components: literature review and synthesis, research goals/problem statement, and study plan. Each component includes two prompts, an initial one and a revised version, with citation formatting applied throughout.

\begin{lstlisting}
LITERATURE REVIEW AND SYNTHESIS:
    Your goal is to expand the literature content of the proposed research idea with more academic context.

    Context: {context}
    Research Idea: {question}
    Web Results (Optional): {web_results}
    
    Instructions for using Literature References:
    1. Critically evaluate the provided academic context and synthesize key insights that are relevant to the research idea.
    2. Identify gaps or limitations in the existing literature that motivate the need for this research.
    3. Explain how the proposed research idea addresses these gaps and contributes to advancing the field.
    
    {CITATION_PROMPT}
    
    Literature Synthesis:
\end{lstlisting}

\begin{lstlisting}
RESEARCH GOALS AND PROBLEM STATEMENTS:
    Your goal is to generate the research goal and corresponding research questions based on the current research idea.
    
    Context: {context}
    Research Idea: {question}
    Literature: {literature}
    
    Instructions for writing research goals: 
    1. Articulate the research goal of this study, such as this study aims to...
    2. Generate corresponding research questions based on the goals with a short descriptive summary.
    3. Consider both theoretical contributions and practical applications.
    4. ONLY formulate 2-3 specific research questions that address gaps in the literature and do not use markdown formatting.
    
    {CITATION_PROMPT}
    
    Research Goals and Problem Statements:
\end{lstlisting}

\begin{lstlisting}
STUDY PLAN:
    You goal is to develop a comprehensive study plan to address research goals.
    
    Context: {context}
    Research Idea: {question}
    Literature: {literature}
    Research Questions: {research_questions}
    
    Instructions for writing study plan: 
    1. Critically use information from the provided context to design an appropriate research plan.
    2. Clearly describe each step for the plan.
    3. Don't repeat Literature and Research Questions and do not use markdown formatting.
    
    {CITATION_PROMPT}
    
    Study Plan:
\end{lstlisting}

\begin{lstlisting}
REVISING:
    You are revising an existing [COMPONENT]. Please carefully revise the previous [COMPONENT] based on the suggested revisions and user instructions and do not use markdown formatting.
    
    Previous Version of [COMPONENT]: {previous_content}
    Suggested Revisions: {evaluation_criteria}
    User Instruction: {user_instruction}
    New Academic Context: {context}
    Web Results (Optional): {web_results}
    
    REVISION GUIDELINES:
    1. SELECTIVE APPROACH:
       a. Analyze the previous version paragraph by paragraph
       b. Try to synthesize information from the new academic context or web results that are useful to address user instruction and suggested revisions 
       b. KEEP sections that are already well-written and accurate
       c. MODIFY ONLY the sections that need improvement
    
    2. COHERENCE:
       a. Ensure smooth transitions between unchanged and revised sections
       b. Maintain logical flow throughout the document
       c. Verify all citations in the previous version are maintained
    
    {CITATION_PROMPT}
    
    [COMPONENT]:
\end{lstlisting}

\begin{lstlisting}
CITATION FORMAT INSTRUCTIONS:
    1. You MUST cite sources using ONLY this format: [CITATION: base_citation_id]
       Example: [CITATION: Smith2020TheoryArti]
    2. Use a separate citation bracket for each source - DO NOT combine multiple citations in one bracket.
       Correct: [CITATION: Smith2020TheoryArti], [CITATION: Jones2021Analysis]
       Incorrect: [CITATION: Smith2020TheoryArti; CITATION: Jones2021Analysis]
    3. DO NOT use numeric citations like [1], [2], etc.
    4. DO NOT use author-date style citations (e.g., "Smith et al. (2020)").
    5. DO NOT use citations from the retrieval context that are not in the missing citations.
    6. Only use citation IDs that appear in the context. DO NOT invent citations.
    7. You may incorporate information from web results if provided, but DO NOT cite web sources.
\end{lstlisting}

\subsection{Prompt Details for Evaluator} \label{appendix:evaluator_prompt}
The \textit{Evaluator} agent provides critical assessment and improvement suggestions for research proposals. It also helps generate new criterion for proposal assessments. We present them as follows.

\begin{lstlisting}
CRITIQUES:
    Evaluate the following research proposal based on the provided criteria.
    Use the retrieved academic context to inform your evaluation where relevant.
    
    # Research Summary
    ## Research Idea
    {research_idea}
    ## Literature Synthesis
    {literature_synthesis}
    ## Research Goals
    {research_goals}
    ## Study Plan
    {study_plan}
    # Evaluation Criteria
    {criteria_text}
    # Retrieved Academic Context (Use this to inform your evaluation)
    {context}
    
    For each criterion, provide one critical reflection:
    Important instructions:
    1. Provide a brief and concise analysis of the research proposal based on the criterion
    2. Reflection could include both strengths and limitations
    3. Ground your evaluation in the retrieved academic context where possible
    4. Do not include citations in your evaluation
    5. Use the exact numeric ID value provided for each criterion, not its name
    Return your evaluation as a JSON object with the following structure:
    {
        "evaluations": [
            {
                "criteriaId": <NUMERIC_ID>,
                "reflections": ["Detailed reflection"],
            },
            ...
        ]
    }
    
    Only return valid JSON. Do not include any other text in your response.
\end{lstlisting}

\begin{lstlisting}
ADDITIONAL CRITIQUES:
    I need additional critical reflections for a research proposal based on a specific criterion.
    
    # Research Summary
    ## Research Idea
    {research_idea}
    ## Literature Synthesis
    {literature_synthesis}
    ## Research Goals
    {research_goals}
    ## Study Plan
    {study_plan}
    # Criterion to Evaluate
    {criterion_name}: {criterion_description}
    # Existing Reflections (DO NOT DUPLICATE THESE)
    {existing_reflections_text}
    # Retrieved Academic Context
    {context}
    
    Please provide 1-2 NEW critical reflections based on the new retrieved academic context that:
    1. Are different from the existing reflections
    2. Provide short and concise analysis of the research proposal based on the criterion
    3. Explore different aspects or angles not covered in existing reflections
    4. Include both strengths and limitations
    5. IMPORTANT, do not specify document references in the new reflections
    Return your reflections as a JSON array:
    {
        "reflections": [
            "Detailed reflection 1",
            "Detailed reflection 2",
            "Detailed reflection 3"
        ]
    }
    
    Only return valid JSON. Do not include any other text in your response.
\end{lstlisting}

\begin{lstlisting}
SUGGESTIONS:
    Generate improvement suggestions for a research proposal based on provided feedback.
    
    # Research Summary
    ## Research Idea: {research_idea}
    ## Literature Synthesis: {literature_synthesis}
    ## Research Goals: {research_goals}
    ## Study Plan: {study_plan}
    # Evaluation Criteria and Feedback
    {criteria_and_feedback}
    Please generate specific improvement suggestions for this research proposal only if you find the criteria are not met:
    1. Prioritize addressing reflections with positive feedback first
    2. Provide concrete, actionable suggestions that directly address the reflections
    3. Include one suggestion per criterion
    4. Are specific enough to implement but not overly prescriptive
    
    Return your suggestions as a JSON object with this structure:
    {
        "improvements": [
            {
                "criteriaId": 1, 
                "criteriaName": "Criterion Name",
                "suggestions": [
                    "Detailed actionable suggestion",
                ]
            },
            ...
        ]
    }
    Only return valid JSON. Do not include any other text in your response.
\end{lstlisting}

\begin{lstlisting}
EVALUATION CRITERIA:
    Generate a new evaluation criteria for assessing research proposals that is different from the existing criteria.
    
    Research idea: {research_idea}
    Existing criteria:
    {criteria_text}
    Create ONE new criteria with:
    1. A short name (1-3 words)
    2. A clear description (1-2 sentences)
    The new criteria should:
    - Not duplicate concepts in existing criteria
    - Be relevant to evaluating academic research
    - Focus on an important aspect not covered by existing criteria
    
    Return your response as a JSON object with the following structure:
    {
        "name": "Criteria Name",
        "description": "Description of the criteria"
    }
    
    Only return valid JSON. Do not include any other text in your response.
\end{lstlisting}

\section{System Evaluation} \label{appendix:system eval}
We conducted two evaluations during system development. First, we tested whether integrating RAG improved generation quality relative to non-RAG generation. Second, we assessed the generation quality against off-the-shelf agentic tools. Both evaluations involved external researchers as pilot participants.

\subsection{Human and LLM Evaluation of RAG Integration} \label{test:RAG-system}
To evaluate whether integrating RAG improves the generation quality (e.g., the content factuality and narrative cohesiveness, compared with the non-RAG version) we used GPT-4o-mini as the LLM service. Two PhD students participated: one from our research team with expertise in creativity and one external researcher specializing in conversational agents (CAs). Each researcher provided field-specific keywords, and the \textit{Ideator} generated five seed ideas. This yielded a total of ten research ideas, listed in Table \ref{tab:evaluation_results}. 

Each seed idea was expanded by the \textit{Writer}, with and without RAG, producing two proposal drafts. To ensure anonymity for evaluation, in-text citations were removed before side-by-side comparison. Three LLM judges (Claude 3.7, DeepSeek-V3, Gemini 2.0) were then prompted to select the better proposal or mark ``undecided'' if neither was preferred. As shown in Table \ref{tab:evaluation_results}, we observed notable differences between LLM and human judgments. Among the ten ideas, LLM judges favored the RAG version in eight cases and disagreed in two. Human researchers preferred the RAG version in four cases, the non-RAG version in two, and were undecided in four. 

Two primary issues were identified. 
First, some RAG-generated proposals over-integrated prior literature. This reduced cohesion across proposal sections. For example, for Idea 2 the external researcher commented: ``Three research questions spanned too many disparate topics, making it hard to see how they could fit into a single study.'' \deleted{Similarly, our internal team member appreciated Idea 3's added specificity from including information literacy and self-efficacy, but felt the second research question on pedagogical strategies was disconnected.}\added{Similarly, another pilot participant noted that the RAG version added specificity, such as in Idea 3, by incorporating concepts like information literacy and measured self-efficacy, but felt that the formulation of research questions were less cohesive: ``These elements were integrated with pedagogical strategies but can't see a clear relationship among them.''} We addressed this issue by refining the prompt to promote more critical use of relevant context and increased the flexibility of the citation checker (instead of forcing all retrieved references). 

The second issue identified was that for undecided cases, researchers found both versions overly general. For example, one noted that neither proposal for idea 6 addressed key challenges, such as tracking cognitive decline in older adults and tailoring the CA design accordingly. When researchers did prefer the RAG version, it was for feasibility: ``I like the first research question of [the RAG version] because it is a question that the science community will be interested in and it is appropriately small and I can imagine concrete steps to operationalize it.'' This second issue appears harder to resolve via prompting alone due to open-ended, domain-specific challenges unresolved in the literature. Ultimately, we decided it would require extensive technical development beyond the scope of our study.

\begin{table*}[t]
\centering
\begin{tabular}{@{}>{\RaggedRight}m{0.3cm}>{\RaggedRight}m{4.5cm}c>{\centering\arraybackslash}m{1.5cm}>{\centering\arraybackslash}m{2cm}>{\centering\arraybackslash}m{1.7cm}>{\centering\arraybackslash}m{2.8cm}@{}}
\toprule
\textbf{ID} & \textbf{Research ideas (abbreviated names)} & \textbf{Type} & \textbf{Claude 3.7} & \textbf{Deepseek-V3} & \textbf{Gemini 2.0} & \textbf{Human evaluator} \\
\midrule
\multirow{2}{*}{1} & \multirow{2}{*}{\parbox{4.5cm}{Personalized search algorithms based on learner interests}} & rag & \checkmark & \checkmark & \checkmark & -- \\
& & non-rag & -- & -- & -- & \checkmark \\
\addlinespace
\multirow{2}{*}{2} & \multirow{2}{*}{\parbox{4.5cm}{Learner interest in information seeking}} & rag & \checkmark & \checkmark & \checkmark & \checkmark \\
& & non-rag & -- & -- & -- & -- \\
\addlinespace
\multirow{2}{*}{3} & \multirow{2}{*}{\parbox{4.5cm}{Curiosity and creativity interplay in AI-collaborative learning}} & rag & -- & \checkmark & -- & -- \\
& & non-rag & \checkmark & -- & \checkmark & \checkmark \\
\addlinespace
\multirow{2}{*}{4} & \multirow{2}{*}{\parbox{4.5cm}{Search approaches that foster epistemic curiosity}} & rag & \checkmark & \checkmark & \checkmark & \checkmark \\
& & non-rag & -- & -- & -- & -- \\
\addlinespace
\multirow{2}{*}{5} & \multirow{2}{*}{\parbox{4.5cm}{AI tailoring learning via curiosity profiles}} & rag & \checkmark & \checkmark & \checkmark & undecided \\
& & non-rag & -- & -- & -- & -- \\
\midrule
\multirow{2}{*}{6} & \multirow{2}{*}{\parbox{4.5cm}{Enhance older adults' decision-making autonomy using CAs}} & rag & \checkmark & \checkmark & \checkmark & undecided \\
& & non-rag & -- & -- & -- & -- \\
\addlinespace
\multirow{2}{*}{7} & \multirow{2}{*}{\parbox{4.5cm}{Create adaptive communication styles of CAs for older adults}} & rag & \checkmark & \checkmark & \checkmark & \checkmark \\
& & non-rag & -- & -- & -- & -- \\
\addlinespace
\multirow{2}{*}{8} & \multirow{2}{*}{\parbox{4.5cm}{Framework for developing the trustworthiness of CAs}} & rag & -- & -- & -- & \checkmark \\
& & non-rag & \checkmark & \checkmark & \checkmark & -- \\
\addlinespace
\multirow{2}{*}{9} & \multirow{2}{*}{\parbox{4.5cm}{Health-related information-seeking using CAs}} & rag & \checkmark & \checkmark & \checkmark & undecided \\
& & non-rag & -- & -- & -- & -- \\
\addlinespace
\multirow{2}{*}{10} & \multirow{2}{*}{\parbox{4.5cm}{Emotional intelligence in CAs to support older adults' autonomy}} & rag & \checkmark & \checkmark & \checkmark & undecided \\
& & non-rag & -- & -- & -- & -- \\
\bottomrule
\end{tabular}
\vspace{0.2em}
\caption{Human and LLM evaluation results of RAG integration. A \checkmark indicates proposals selected by either LLM judges or human evaluators, while `–' indicates those not selected.}
\Description{Human and LLM evaluation results of RAG integration.}
\label{tab:evaluation_results}
\end{table*}

\begin{table*}[t]
\centering
\begin{tabular}{@{}>{\RaggedRight}p{0.3cm}>{\RaggedRight}m{5cm}>{\centering\arraybackslash}m{1.5cm}>{\centering\arraybackslash}m{1.5cm}>{\centering\arraybackslash}m{1.5cm}>{\centering\arraybackslash}m{1.5cm}@{}}
\toprule
\textbf{ID} & \textbf{Research Idea \newline (abbreviated names)} & \textbf{Deep Research} & \textbf{ChatGPT (GPT-4.1 + Search)} & \textbf{Gemini (2.0 Flash + Search)} & \textbf{Ours (GPT-4.1)} \\
\midrule
11 & Closed-loop PFAS receptor discovery via Bayesian optimization for scalable water remediation & 1 & 2 & 4 & 3 \\
\addlinespace[0.3em]
12 & Quantum-inspired RL with physics-informed NN for topological phonon transport in cryogenic materials & 1 & 3 & 4 & 2 \\
\addlinespace[0.3em]
13 & Interactive vs. traditional storytelling for fostering children's empathy and emotional intelligence & 1 & 3 & 4 & 2 \\
\bottomrule
\end{tabular}
\vspace{0.2em}
\caption{Pilot evaluation rankings for three research proposals across four LLM-based systems (lower rank indicates better proposal quality). The off-the-shelf LLM tools, publicly available at the time, were powered by their respective versions as of early June 2025. Our system ranked second or third, demonstrating a sufficiently competitive performance for real user test.}
\Description{Pilot evaluation rankings for three research proposals across four LLM-based systems}
\label{tab:pilot_evaluation}
\end{table*}

\subsection{Expert Assessment of Agentic System} \label{test:agentic-system}

We conducted pilot testing with three additional researchers: two PhD students (Chemistry and Child Development) and a postdoctoral fellow in Physics. The goal was to evaluate proposal quality across domains beyond our HCI and NLP expertise and to compare our system with established agentic tools. 
For this evaluation, we employed GPT-4.1 as the LLM for our ideation agentic system, offering the best performance and fastest response time available. This yielded three new ideas 11-13 (Table \ref{tab:pilot_evaluation}). 

Each participant then generated proposals based on their respective idea using \textit{Deep Research}, standard \textit{ChatGPT}, and \textit{Gemini}. To compare generated proposals across the three different LLMs, we used the initial drafts from each without any further revision. Participants then ranked all drafts from best to worse based on three criteria: Novelty, Feasibility, and Impact (the same three default criteria used for the \textit{Evaluator}). The evaluation goal was not to show superior performance but to assess whether our system met a reasonable standard for use in comparison to commonly used research tools (i.e., that 
the system was sufficiently competitive for a formal user study).

As shown in Table \ref{tab:pilot_evaluation}, our system ranked second or third, evidencing comparable performance at the time of testing. However, researcher feedback revealed similar cohesion issues noted in Section \ref{test:agentic-system}. For example, for Idea 11, researchers commented that ``It is unclear whether the study goal is elucidating the mechanisms of toxicity or developing PFAS capture strategies in the introduction.'' In comparison, the proposal generated by Deep Research was described as more ``well-defined'' and ``thorough.''

\section{Study Protocol: Moderation Notes, Task Description, and Interview Questions} \label{appendix:study protocol}

\subsection{Moderation Notes}

\textit{Ice-breaking Conversations:} Before we jump into using our tool and asking you to generate a research proposal, we’d like to start with a few quick questions about your experience using AI tools in your research. What research project are you currently working on? Based on your survey response, you mentioned using. Can you tell us a bit about how you’ve used those tools in your research, and how helpful they were?

\vspace{0.1cm}

\noindent \textit{Tool Introduction:} Great, now let’s move on to introducing the tool. We’re building an AI tool designed to help researchers like you generate research ideas more effectively. What you’ll be using today is an early version of that tool. We’ve prepared a short video guide that explains the tool’s features and how they work. I’ll share it through Zoom chat. [Share the link] Please take a few minutes to watch it. Let us know when you’re done. This is also our text version of the tool introduction, you can open it anytime and refer to it. 

\subsection{Task Description}
Now, please imagine you’re preparing a research proposal for a project that you are thinking about conducting, using your own research topics. You’ll be using this AI tool to help you draft ideas. Your goal is to come up with as many research ideas as you’d like within your area of interest. Each time you feel satisfied with a draft, please save it. After using the tool, we’ll ask you to choose the one draft you think is the best among the ones you saved. You’ll have 30 minutes and we will let you know when 3 minutes left and time’s up.

\subsection{Interview Questions}
\label{appendix:interview questions}

\begin{itemize}
    \item Tool Experience
    \begin{itemize}
        \item Let’s start with your overall experience using the tool. Could you briefly go over how you used the tool? 
        \item Let’s start talking about idea generation. Do you think the tool helped, or did it make it harder to come up with research ideas? (Follow-up: What made you feel that way? Could you tell me a bit more?)
        \item The tool includes several features, [remind the features the tool have, if participant cannot come up with features: like suggesting keywords, helping your literature search, synthesizing papers you selected, generating research goals and study plans automatically, and having another AI give feedback on your drafts] Which of these features felt the effortful (hardest) to use? On the flip side, which feature felt the most rewarding (helpful) to use?
        \item Actually, our tool has three different versions, each representing a different level of human involvement. Some features are intentionally disabled in the Daisy and Rose versions. Ask questions:
        \begin{itemize}
            \item How do you think the available or missing features as shown would influence your ability to generate strong research ideas?
            \item Will the availability of those features also affect your sense of control or ownership in generating ideas?
        \end{itemize}
        \item Now, could you please open the drafts you saved using the tool and drop them in the Zoom chat? They should be in your folder. [Wait for the drafts to be shared.] Take a moment to review them, and pick the one you feel is your best draft. Let me know when you’ve made your choice. [When they choose it] What made you feel this was your best draft? 
        \item Overall, how would you rate the quality of the AI-generated content? Does it generally seem high-quality, or does it feel like it’s just making things up?
    \end{itemize}
    \item Trade-offs of Human–AI Co‑Developing Research Proposals
    \begin{itemize}
        \item Optional, ask when observed: Why did you decide to use the “instruct LLM to revise” function  often rather than editing the content directly yourself? 
        \item Optional, ask when observed: I noticed that you rarely use the “use LLM to evaluate” function. Could you share why?
        \item I have a quick question about one of your final survey responses. There was a question asking how you viewed the authorship of your proposal drafts. You selected [read participant’s choice]. 
        \begin{itemize}
            \item Could you explain why you feel the proposal is primarily [ownership type]?
            \item What aspects of proposal writing might make you feel that?
        \end{itemize}
        \item Given that our tool relies heavily on AI, do you think the human role in research proposal writing is shifting from mainly writing to more reviewing and evaluating AI-generated content?
        \begin{itemize}
            \item If so, how do you feel about this change? Do you see it as a positive or negative development? [follow-up: What qualities or skills do you think are important for someone to be a good evaluator in this kind of workflow?]
            \item If not, how would you distinguish the role of a writer from that of an evaluator?
        \end{itemize}
    \end{itemize}
    \item Role of Human Effort and Value in AI‑Assisted Research
    \begin{itemize}
        \item Thinking broadly about your overall research process, including brainstorming, reading, planning, and writing, which parts do you find the most challenging?
        \item Which parts do you find the most enjoyable to do yourself? (Follow-up: What makes these parts meaningful for you?)
        \item Which parts of this process do you think AI could reasonably take over? Or, are there any aspects you feel it should not be used for?
        \item Optional: Last but not least, do you have any suggestions for how we can design AI tools to better support researchers like you?
        \item Optional: Are there any research activities unique to your discipline that you haven’t yet mentioned? How do you envision AI supporting or enhancing those?
    \end{itemize}
\end{itemize}

\section{Representative Use Case} \label{appendix:representative case}
\added{To illustrate how participants used the drafting time effectively, we present a representative case from one user, P10 (Intensive, PhD student). Due to university IRB restrictions, we were not permitted to share full document snapshots or detailed participant edits publicly. Therefore, we extracted partial interview excerpts, user prompts, and key edits, and aligned them with AI actions to construct a step-by-step case description. This aims to clearly communicate the interplay between AI-generated and participant-authored contributions.} 

\added{P10 spent approximately 40 minutes expanding two seed ideas into complete drafts. Below, we describe how P10 elaborated on the first idea.}

\begin{enumerate}
\item \textbf{Exploring the initial idea and specifying literature ($\sim$3 minutes)}
    \begin{itemize}
        \item \textbf{Human:} Entered an initial research idea: ``sentiment analysis, toxicity analysis, open source software security.''
        \item \textbf{AI:} Returned four seed ideas. 
         \item \textbf{Human:} Selected and edited one seed idea: ``Create a real-time dashboard that tracks developer sentiment and toxicity in open source projects, correlating these trends with security vulnerabilities to support proactive interventions."
        \item \textbf{Human:} Opened the Search Sidebar and typed keywords to search paper, including ``open source software, toxicity analysis, sentiment analysis.''
        \item \textbf{AI:} Returned 100 relevant results.
        \item \textbf{Human:} Reviewed paper snippets, opened and read papers (e.g., ``analyzing and detecting toxicities in developer online chatrooms''), and selected 8 papers in total for inclusion.
    \end{itemize}
\item \textbf{Generating and revising the literature review ($\sim$5 minutes)}
    \begin{itemize}
        \item \textbf{AI:} Generated the \textit{Literature Exploration and Synthesis} section based on the selected papers.
        \item \textbf{Human:} Reviewed the generated section and used feature \textit{Instruct LLM to revise}: ``...add more technical details, such as impact of sentiment and toxicity to contributor health.''
        \item \textbf{AI:} Regenerated the literature review based on the instruction.
        \item \textbf{Human:} Reviewed the revision and triggered \textit{Generate full proposal}, spanning literature review, research goals, and study plan.
    \end{itemize}
\item \textbf{Refining research goals and questions ($\sim$3 minutes)}
    \begin{itemize}
        \item \textbf{AI:} Generated the \textit{Research Goals} section with two initial research questions relating sentiment, toxicity, and security vulnerabilities in OSS.
        \item \textbf{Human:} Used detailed prompting: ``...add more research questions and sub-questions that showcase a road map. Think like an esteemed computer science researcher.''
        \item \textbf{AI:} Produced a new version with three research questions, each with two sub-questions.
        \item \textbf{Human:} Used \textit{Version\_comparison} and selected the revised version.
    \end{itemize}
\item \textbf{Evaluating agent critiques and providing feedback ($\sim$3 minutes)}
    \begin{itemize}
        \item \textbf{Human:} Triggered \textit{Use AI for evaluation and revision}.
        \item \textbf{AI:} Evaluated novelty, feasibility, and impact of the draft.
        \item \textbf{Human:} Read impact feedback (e.g., ``need for more detail on intervention strategies and sensitivity to project characteristics'') and requested \textit{Generate more reflections}.
        \item \textbf{AI:} Generated two additional impact-focused reflections.
        \item \textbf{Human:} Used thumps-up features to support three AI reviews i.e., ``...strengthened by more specific details on intervention strategies...,'' ``...advance prevention and intervention techniques for open source security...,'' and ``...significant practical impact in creating healthier OSS ecosystems...''
        \item \textbf{Human:} Triggered \textit{Revise based on AI reflections} and clicked \textit{Refine Research Idea}.
    \end{itemize}
\item \textbf{Regenerating the full draft and saving outputs ($\sim$3 minutes)}
    \begin{itemize}
        \item \textbf{AI:} Regenerated all proposal sections and expanded the literature with six new papers based on accepted feedback.
        \item \textbf{Human:} Reviewed the updated draft and downloaded it as the first completed output.
    \end{itemize}
\end{enumerate}

\section{Participant Demographics and Research Domains} \label{appendix:demographics}




\begin{table*}[t]  
\centering
\resizebox{\linewidth}{!}{%
\begin{tabular}{@{}p{0.03\linewidth}>{\RaggedRight}p{0.17\linewidth}>{\RaggedRight}p{0.08\linewidth}>{\RaggedRight}p{0.25\linewidth}>{\RaggedRight\arraybackslash}p{0.58\linewidth}@{}}
\toprule
\textbf{ID} & \textbf{Position \& Sex} & \textbf{Location} & \textbf{Research Domain} & \textbf{Research Subjects} \\
\midrule
P01 & PhD-Student (F) & US & Chemistry & Data-driven molecules design \\
P02 & PhD-Student (F)& US & Medicine & Cancer risk factors and prognosis \\
P03 & PhD-Student (F) & US & Urban Planning & New path development in old industrial regions \\
P04 & Research Scientist (F) & Jamaica & Education & Self-regulated and instruction learning \\
P05 & PhD-Student (F) & US & Information Science & Interdisciplinarity/transdisciplinarity sustainable development goals research \\
P06 & PhD-Student (F) & US & Biomedical Engineering & RNA-based diagnostics and therapeutics \\
P07 & PhD-Student (M) & US & Materials Science & Soft polymer materials and hydrogels \\
P08 & Postdoc (F) & France & Epidemiology & Characterization of antenatal corticosteroid recipients for preterm birth intervention \\
P09 & PhD-Student (F) & US & Human-computer Interaction & Ecosystem-level human-robot interaction \\
P10 & PhD-Student (M) & US & Computer Science & Open source security and NLP \\
P11 & Professor (F) & Korea & Media and Communication & Mis/Disinformation \\
P12 & Research Scientist (M) & US & Computer Science & Human-centered research software engineering for high-performance systems \\
P13 & PhD-Student (F) & US & Information Science & Collective behavior in knowledge-sharing online communities \\
P14 & Postdoc (M) & US & Computational Pathology & Guided diffusion models for cross-modality style transfer and cell type classification in 3D spatial omics \\
P15 & PhD-Student (F) & US & Political Science & Geopolitics and threat perception in foreign policy \\
P16 & PhD-Student (M) & US & Computer Science & Diffusion language models, continual learning, and preference optimization \\
P17 & PhD-Student (M) & US & Physics & Quantum field theory, string theory, and holography \\
P18 & Professor (M) & US & Information Science & Data governance/management, data/process quality, and latent reasoning techniques \\
P19 & PhD-Student (M) & US & Media and Communication & Digital religion \\
P20 & PhD-Student (F) & Canada & Information Science & Scholarly communication and research data management \\
P21 & PhD-Student (M) & US & Information Science & Information retrieval \\
P22 & PhD-Student (M) & US & Computer Science & Vision language action models for robotic task planning \\
P23 & Research Scientist (M) & US & Aerospace Engineering & Surrogate, foundation modeling, and time series prediction in fluid dynamics \\
P24 & PhD-Student (M) & US & Electrical and Computer Engineering & Reliability of MRI reconstruction algorithms \\
P25 & PhD-Student (F) & US & Education & Collaborative learning, cognitive processes, and behavioral intentions \\
P26 & PhD-Student (M) & US & Mechanical Engineering & Visual localization \\
P27 & Postdoc (F) & Germany & Neurobiology & Biology function of DExH-box helicase 9 \\
P28 & Postdoc (F) & US & Information Science & Cognitive impacts of AI systems \\
P29 & PhD-Student (F) & US & Sociology & Mental health, adolescent, and social network \\
P30 & PhD-Student (M) & US & Computer Science & Consistency and robustness of large language model \\
P31 & PhD-Student (M) & India & Information Science & Performance ranking on search engine, citation dynamics, and knowledge graph for language models \\
P32 & PhD-Student (M) & US & Chemistry & Atomistic scale simulation, machine learning interatomic potential, and solid state electrolyte \\
P33 & Professor (F) & US & Art and Design & Civic design for healthcare \\
P34 & PhD-Student (M) & US & Business & Organizational isomorphism and AI in Organizations \\
P35 & PhD-Student (M) & US & Computer Science & Gesture elicitation studies in AR environment \\
P36 & PhD-Student (F) & US & Political Science & International relations, foreign direct investment, and military aid in conflict \\
P37 & PhD-Student (F) & US & Political Science & Conflict, peacebuilding, and civil society \\
P38 & PhD-Student (M) & US & Computer Science & Misinformation detection and fairness in large language models \\
P39 & PhD-Student (F) & US & Human-computer Interaction & Social cognition and AI companionship \\
P40 & PhD-Student (F) & US & Education & Digital game-based learning and physiological computing \\
P41 & PhD-Student (F) & US & Nutritional Science & Prostate cancer combinatorial drug therapy \\
P42 & Postdoc (F) & US & Information Science & Privacy \\
P43 & PhD-Student (M) & US & Computer Science & Reinforcement learning for robotics \\
P44 & PhD-Student (M) & US & Political Science & Political biases in large language models \\
P45 & PhD-Student (F) & US & Education & Instructional systems technology \\
P46 & PhD-Student (F) & US & Political Science & Corruption in public procurement \\
P47 & Postdoc (F) & US & Human-computer Interaction & Decision support tools for healthcare \\
P48 & PhD-Student (M) & US & Chemistry & Pulmonary drug delivery, gene therapy, and polymeric nanoparticles \\
P49 & PhD-Student (M) & US & Chemistry & Density functional theory, machine learning force fields, and model optimization \\
P50 & Research Scientist (F) & US & Information Science & Misinformation interventions and academic integrity policy in education \\
P51 & PhD-Student (M) & US & Biology & Sorghum breeding and genetics, improved grain yield with biotech \\
P52 & PhD-Student (F) & Germany & Sinology & Literature motif in Chinese Wuxia \\
P53 & Postdoc (M) & US & Physics & Multiscale thermal transport and quantum materials \\
P54 & PhD-Student (F) & US & Information Science & Search as learning with voice assistants \\
\bottomrule
\end{tabular}
}
\caption{Participant Demographics and Research Domains}
\label{tab:demographics}
\end{table*}

\section{Additional Statistical Results} \label{appendix:extra_statistics}

In this section, we present additional statistical analyses to supplement the quantitative results reported in Section~\ref{subsec:quantitative_results}.

\subsection{Post-hoc Analysis for Behavioral Metrics} \label{appendix:post-hoc}

We compute Tukey's HSD post-hoc analysis for behavioral metrics that show statistically significance from the one-way ANOVA test reported in Section \ref{quant:behavioral}, including \textit{prompt\_full\_proposal} and \textit{select\_improvements}.

\begin{table*}[t]
\centering
\begin{tabular}{@{}lllllll@{}}
\toprule
\textbf{Group 1} & \textbf{Group 2} & \textbf{Mean Diff.} & \textbf{p-value} & \textbf{Lower} & \textbf{Upper} & \textbf{Reject} \\
\midrule
Intensive & Medium  & 0.6672  & 0.9314 & -3.8246 & 5.1590  & False \\
Intensive & Low & 6.1516  & 0.0049 & 1.6598  & 10.6434 & True  \\
Medium    & Low & 5.4844  & 0.0181 & 0.8040  & 10.1647 & True  \\
\bottomrule
\end{tabular}
\vspace{0.2cm}
\caption{Tukey's HSD post-hoc analysis for \textit{Prompt\_full\_proposal}.}
\label{tab:prompt_full_proposal}
\end{table*}

As shown in Table \ref{tab:prompt_full_proposal}, participants in the low control-level prompted full proposals significantly more often than those in the Intensive ($M_\text{diff}=6.15$, $p=0.0049$) and Medium ($M_\text{diff}=5.48$, $p=0.018$) conditions. There was no significant difference between the Intensive and Medium conditions. This indicates that the low control encourages greater usage on the system to generate full proposals.

\begin{table*}[t]
\centering
\begin{tabular}{@{}lllllll@{}}
\toprule
\textbf{Group 1} & \textbf{Group 2} & \textbf{Mean Diff.} & \textbf{p-value} & \textbf{Lower} & \textbf{Upper} & \textbf{Reject} \\
\midrule
Intensive & Medium  & 1.2616  & 0.5931 & -1.8574 & 4.3759  & False \\
Intensive & Low & -2.6063 & 0.1176 & -5.7195 & 0.5090  & False \\
Medium    & Low & -3.8669 & 0.0160 & -7.1119 & -0.6219 & True  \\
\bottomrule
\end{tabular}
\vspace{0.2cm}
\caption{Tukey's HSD post-hoc analysis for \textit{Select\_improvements}.}
\label{tab:select_improvements}
\end{table*}

In table~\ref{tab:select_improvements} for \textit{select\_improvements}, we found that participants in the Medium condition selected significantly more improvements than those in the Low condition ($M_\text{diff}=-3.87$, $p=0.016$). No significant differences were observed between the Intensive condition and the other two conditions. This suggests that the Medium control level encourages greater engagement in iterative refinement compared with the Low control.

\subsection{Fine-grained Analysis for the Trade-offs between Individual Creativity Support and Effort Dimension} \label{appendix:fine-grained_tradeoffs}

\begin{figure*}[t]
    \centering
    \includegraphics[width=\textwidth]{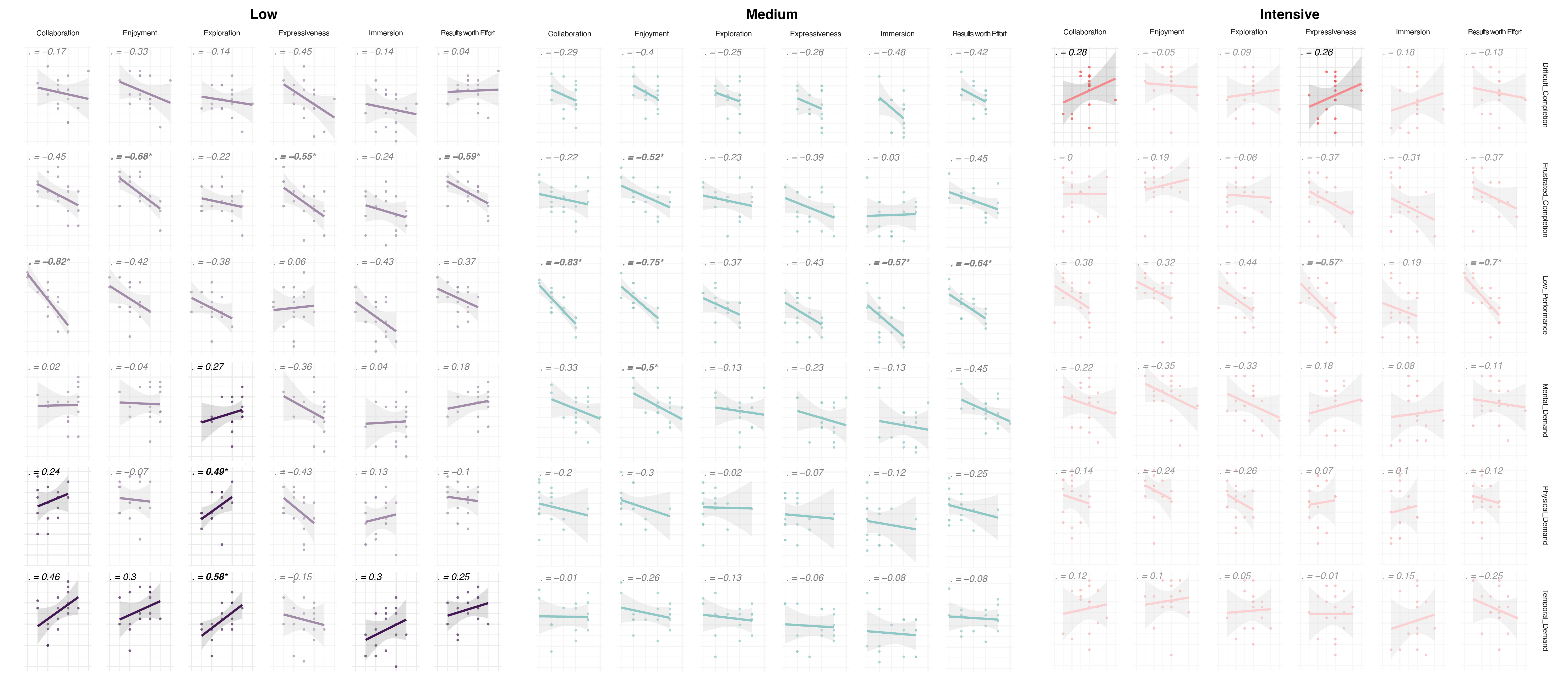}
    \caption{Correlations between each specific creativity dimension (unweighted scores) and effort dimension across three control levels. Creativity support dimensions are shown on the y-axis and effort scores on the x-axis. Pearson correlations were applied due to the parametric nature of both variables. As most correlations were negative, we highlight positive trends (above 0.2) to illustrate distinct patterns. Statistically significant correlations are shown in bold.}
    \label{fig:creativity_effort_tradeoff_grid}
\end{figure*}

To analyze the relationship between specific creativity and effort dimensions across the three control levels, we primarily used unweighted creativity scores. As noted by \citet{Cherry2014-xs}, weighted scores are most useful when comparing the relative importance of different creativity support dimensions. However, because our focus was not on comparing dimensions against each other but rather on examining how each creativity dimension relates to specific types of effort, unweighted scores provided a more objective and accurate measure for this analysis.

As shown in Figure \ref{fig:creativity_effort_tradeoff_grid}, increasing effort generally corresponded to decreases in several creativity dimensions. However, nuanced variations exist in different control levels (see highlighted subgraphs). 

At the low-control level, statistically significant negative correlations were observed for enjoyment ($R = -0.98$, $p < .05$), expressiveness ($R = -0.55$, $p < .05$), and value ($R = -0.59$, $p < .05$) with increasing task difficulty, and for collaboration ($R = -0.82$, $p < .05$) with increasing frustration. However, certain aspects of creativity support increased with higher effort. For example, exploration showed a positive relationship with mental demand ($R = 0.27$, $p > .05$) and physical demand ($R = -0.49$, $p < .05$). Additionally, collaboration ($R = 0.46$, $p > .05$), exploration ($R = 0.58$, $p < .05$), immersion ($R = 0.30$, $p > .05$), and value ($R = 0.25$, $p > .05$) increased with temporal demand.

At Medium-control level, increases in various types of effort were associated with decreases in all creativity dimensions. In particular, enjoyment decreased significantly with increasing frustration ($R = -0.52$, $p < .05$). Collaboration ($R = -0.83$, $p < .05$), enjoyment ($R = -0.75$, $p < .05$), immersion ($R = -0.57$, $p < .05$), and value ($R = -0.64$, $p < .05$) all significantly decreased with increasing lower performance. Additionally, enjoyment significantly decreased with higher mental demand ($R = -0.50$, $p < .05$).

At Intensive-control level, certain aspects of creativity support significant decreases in some types of effort. For example, expressiveness ($R = 0.57$, $p < .05$) and value ($R = 0.70$, $p < .05$) decrease with increasing lower performance. In contrast, some creativity support dimensions tended to increase as task difficulty increased such as collaboration ($R = 0.28$, $p > .05$) and expressiveness ($R = 0.26$, $p > .05$), although these correlations were not statistically significant.

\end{document}